\documentclass[twocolumn,pra,aps,amsmath,amssymb,superscriptaddress,showpacs]{revtex4-1}

\renewcommand{\vec}[1]{{\mathbf #1}}
\renewcommand{\bf}[1]{{\normalfont \bfseries #1}}

\DeclareMathOperator{\Tr}{Tr}
\DeclareMathOperator*{\SumInt}{
\mathchoice
  {\ooalign{$\displaystyle\sum$\cr\hidewidth$\displaystyle\int$\hidewidth\cr}}
  {\ooalign{\raisebox{.14\height}{\scalebox{.7}{$\textstyle\sum$}}\cr\hidewidth$\textstyle\int$\hidewidth\cr}}
  {\ooalign{\raisebox{.2\height}{\scalebox{.6}{$\scriptstyle\sum$}}\cr$\scriptstyle\int$\cr}}
  {\ooalign{\raisebox{.2\height}{\scalebox{.6}{$\scriptstyle\sum$}}\cr$\scriptstyle\int$\cr}}
}

\usepackage[english]{babel}
\usepackage{mathrsfs}     
\usepackage{graphicx}
\usepackage{bbm}
\usepackage[usenames,dvipsnames]{color}
\usepackage{natbib}
\usepackage{url}
\usepackage[bookmarks=true]{hyperref}
\usepackage{commath} 
\hypersetup{colorlinks=true, linkcolor=blue, urlcolor=blue, citecolor=blue} 
\usepackage{xfrac} 
\usepackage{braket}
\usepackage{acronym}
\usepackage{setspace}

\usepackage{latexsym,amsfonts,amsmath,amssymb}
\usepackage{graphics,graphicx,color}
\usepackage[normalem]{ulem}

\begin{document}

\title{
Tailored pump-probe transient spectroscopy with time-dependent
density-functional theory: controlling absorption spectra
}

\author{Jessica Walkenhorst}
\affiliation{Nano-Bio Spectroscopy Group and ETSF Scientific Development Center, Departamento de Quimica, 
Universidad del Pa\'is Vasco UPV/EHU, Avenida de 
Tolosa 72, E-20018, San Sebasti\'an, Spain}\email[]{walkenho@gmail.com}
\author{Umberto De~Giovannini}\email[]{umberto.degiovannini@ehu.es}
\affiliation{Nano-Bio Spectroscopy Group and ETSF Scientific Development Center, Departamento de Quimica, 
Universidad del Pa\'is Vasco UPV/EHU, Avenida de 
Tolosa 72, E-20018, San Sebasti\'an, Spain}
\author{Alberto Castro}\email[]{acastro@bifi.es}
\affiliation{ARAID Foundation - Institute for Biocomputation
and Physics of Complex Systems, University of Zaragoza Mariano Esquillor Gómez s/n, 50018 Zaragoza, (Spain)}
\author{Angel Rubio}\email[]{angel.rubio@mpsd.mpg.de}
\affiliation{Nano-Bio Spectroscopy Group and ETSF Scientific Development Center, Departamento de Quimica, 
Universidad del Pa\'is Vasco UPV/EHU, Avenida de 
Tolosa 72, E-20018, San Sebasti\'an, Spain}
\affiliation{Max Planck Institute for the Structure and Dynamics of Matter
  Hamburg, Germany}

\begin{abstract}
Recent advances in laser technology allow us
  to follow electronic motion at its natural time-scale with ultra-fast time resolution,
  leading the way towards attosecond physics experiments of extreme precision.
  In this work, we assess the use of
  tailored pumps in order to enhance (or reduce) some given features of the
  probe absorption (for example, absorption in the visible range of otherwise
  transparent samples). This type of manipulation of the system response could
  be helpful for its full characterization, since it would allow to visualize
  transitions that are dark when using unshaped pulses. In order to
  investigate these possibilities, we perform first a theoretical analysis of the
  non-equilibrium response function in this context, aided by one simple
  numerical model of the Hydrogen atom. Then, we proceed to investigate the
  feasibility of using time-dependent density-functional theory as a means to
  implement, theoretically, this absorption-optimization idea, for more
  complex atoms or molecules. 
  We conclude that the proposed idea could in principle be brought to the laboratory: tailored pump pulses 
  can excite systems into light-absorbing states. However, 
  we also highlight the severe numerical and theoretical difficulties posed by the problem: 
  large-scale non-equilibrium quantum dynamics are cumbersome, even with TDDFT, 
  and the shortcomings of state-of-the-art TDDFT functionals 
  may still be serious for these out-of-equilibrium situations.
\end{abstract}

\date{\today}

\maketitle

\section{Introduction} %
\label{sec:introduction}

Time-resolved pump-probe experiments are powerful
techniques to study the dynamics of atoms and molecules: 
the pump pulse triggers the dynamics, which is then monitored by measuring the time-dependent 
response of the excited system to a probe pulse. 
The time-resolution of this technique has increased over the years, and 
nowadays, it can be used to observe the electron dynamics in real time, giving rise to the field of 
attosecond physics \cite{krausz-2009, scrinzi-2006}.

A suitable setup to observe charge-neutral excitations is the time-resolved photoabsorption or transient absorption
spectroscopy (TAS), where the time-dependent optical absorption of the probe is measured. 
TAS can of course be used to look at longer time resolutions:
if we look at molecular reaction on the
scale of tens or hundreds of femtoseconds, the atomic structure will
have time to re-arrange. These techniques are thus mainly
employed in femtochemistry \cite{book-zewail, zewail-2000} 
to observe and control modification, creation, or destruction of bonds.
TAS has been successfully employed, for example, to watch the
first photo-synthetic events in cholorophylls and
carotenoids \cite{berera2009}
(a review describing
the essentials of this technique can be found in
Ref.~\cite{foggi2001}).  
If, on the contrary, one wants to study the electronic dynamics \emph{only}, 
disentangling it from the vibronic degrees of freedom,
then one must perform TAS with attosecond pulses \cite{pfeifer2008}, 
a possibility recently demonstrated \cite{goulielmakis-2010, holler2011}.

The theoretical description of these processes, which involve non-linear
light-matter interaction, and the ensuing non-equilibrium electron dynamics, is
challenging. Time-dependent density functional theory (TDDFT) \cite{runge-1984, book-tddft,tddft-specialissue} 
is a well-established tool
to compute the response of a many-electron system to arbitrary perturbations.
Traditionally, the vast majority of TDDFT applications have addressed the first-order
response of the ground-state system to weak electric fields -- which
can provide the absorption spectrum, the optically-allowed excitation
energies and oscillator strengths, etc. Nevertheless, the extension of TDDFT to the 
description of excited state spectral properties and its ability to simulate
transient absorption spectroscopy (TAS) 
has recently been demonstrated \cite{degiovannini-2012, degiovannini-2013}.

In this work, we are not only interested in simulating attosecond
TAS of atoms and molecules, but in studying the possibility of tailoring the
pump to control the spectra. In fact, the measurement and control of ultrafast processes are inherently intertwined:
quantum optimal control theory (QOCT)~\cite{Brif2010, werschnik-2007} can be viewed as the
\emph{inverse} of theoretical spectroscopy: rather than attempting to predict the
reaction of a quantum system to a perturbation, it attempts to find the
perturbation that induces a given reaction in a given quantum system. It is
the quantum version of the more general control theory 
\cite{Shi1988, peirce-1988, kosloff-1989, Luenberger1969,Luenberger1979}, which was needed
given the fast advances in experimental quantum 
control~\cite{Brumer1986541,Brumer1986177,shapiro-2003,Tannor1985541,Herek199415, Gaubatz1988463, 
doi:10.1146/annurev.physchem.59.032607.093818,judson-1992,bardeen-1997,assion-1998}.

The possibility of combining QOCT with TDDFT has been established recently~\cite{castro-2012}. Furthermore it has been
shown, that it can be used to optimize strong-field ionization \cite{hellgren-2013}, photo-induced dissociation
\cite{castro-2013} and is compatible with Ehrenfest dynamics \cite{castro-2014}. Very recently, Krieger \emph{et al}.\
used TDDFT to study the influence of laser intensity, frequency and duration on the laser-induced 
demagnetization process in bulk materials, which takes place on time scales of $<$ 20 fs \cite{krieger-2015}. 

Here, we take the first steps towards the use of this combination of TDDFT with QOCT
to control excited state spectra of finite systems. This idea is very much related to the
concept of electromagnetically induced transparency \cite{harris-1990, fleischhauer-2005}.
Control of the absorption spectra may mean its elimination or reduction, or
its increase.
Our \emph{gedanken} setup throughout this paper is the following: for a
certain time interval $[0,T]$ a quantum system is driven by a ``classical'' pump pulse
$\mathscr{E}(t)$ whose precise shape can be manipulated. After the pump has ended, the (linear) 
response of the system to some later perturbation is calculated. 
Our goal is to design the shape of the pump pulse in such a way, that the response 
to some later perturbation is \emph{optimal} in some given way. 
In particular, we demonstrate how the tailored pump pulses may be used to
transform a transparent atom or molecule into an excited one that absorbs in
the visible.

This paper is structured as follows. In Section~\ref{sec:spec_ex}, we analyze 
the optical linear response of a system in an excited state, looking at the location 
and the shape of the resulting spectral peaks and their time-dependence. In Section~\ref{sec:qoct} 
we present the theory of quantum optimal control and how it can be applied to optimize spectral 
properties of systems in excited states. 
We then bring these concepts into application in Section~\ref{sec:applications}. In Section~\ref{sec:hydrogen}
we illustrate the conclusions gained in Section~\ref{sec:spec_ex} using the analyticly solvable Hydrogen atom.
We then proceed to demonstrate control of the excited state properties in this system. 
In Section~\ref{sec:tddft} we finally combine our methodology with
time-dependent density functional theory, 
first for the Helium atom, and then for the Methane molecule.
We conclude our work in Section~\ref{sec:conclusions}.

\section{Short Review of Out-of-equilibrium (Pumped) Absorption Spectra}\label{sec:spec_ex} %

In pump-probe spectroscopy, the probe may arrive after, during, or even before
the pump; in this work, we consider a non-overlapping regime in which the
probe arrives after the pump has vanished.
The time evolution of a system after the end of the pump is described by the
Hamiltonian (atomic units will be used hereafter):
\begin{equation}
    \hat H(t)=\mathscr{\hat H} + F(t) \hat D_\mu \, ,
    \label{eq:hamiltonian_nonoverlapping}
\end{equation}
where $\mathscr{\hat H}$ is the static Hamiltonian, that describes the system itself
and $F(t) \hat D_\mu$ is the coupling to a probe pulse
via the dipole operator
\begin{equation}
    \hat D_{\mu}=-\sum_{i=1}^N \hat r_{\mu}^{(i)}
    \label{eq:dipole_op}
\end{equation}
in which $N$ is the number of electrons in the system, and $\mu=x,y,z$
determines the polarization direction.
Note, that the implementation of the coupling via the dipole operator is an approximation and could be removed 
in practical implementations since the theory handles non-dipolar fields. 
Also note, that we work in the length gauge all over the paper.

If, at the time $t=T$, the system has been driven by the previous pump $\mathscr{E}$ to the state
$\vert\Psi\left[\mathscr{E}\right](T)\rangle$, the complete dipole-dipole response function for the perturbation at
later times is given by:%
\begin{eqnarray} \label{eq:retarded_response_function_recast}
\chi_{\hat{D}_\mu,\hat{D}_\nu}&&\left[\mathscr{E}\right](t,t') = \\
&& -i\theta(t-t')
\langle\Psi\left[\mathscr{E}\right](T)\vert
\left[
\hat{D}_{\mu I}(t),\hat{D}_{\nu I}(t')
\right]
\vert\Psi\left[\mathscr{E}\right](T)\rangle\, ,\nonumber   
\end{eqnarray}
where the operators are expressed in the interaction picture.

The difference between Eq.~\eqref{eq:retarded_response_function_recast} and
the equilibrium response function \cite{book-fetter} or the response function of a system in 
a many-body eigenstate is that the two times 
$t$ and $t'$ cannot be reduced to only one by making
use of the time-translational invariance.
Since $\vert\Psi\left[\mathscr{E}\right](T)\rangle$ depends on the pump, 
the first-order response of the system does explicitly depend on both the
pump $\mathscr{E}$ and the probe $F$. It is given by:
\begin{equation}
D^{(1)}_{\mu\nu}\left[\mathscr{E}, F\right](t) = \int_{T}^{t}\!{\rm d}t' F(t')\chi_{\hat{D}_\mu,\hat{D}_\nu}\left[\mathscr{E}\right](t,t')\,.
    \label{eq:dipole_response}
\end{equation}
An intuitive physical meaning can be gained from this equation for the
response function: it is the first order of the system response, if we consider
a sudden perturbation at $t'=T+\tau: F(t) = \delta(t-(T+\tau))$, where $\tau$ denotes the 
delay between the end of the pump and the perturbation:
\begin{equation}
    \chi_{\hat{D}_\mu,\hat{D}_\nu}\left[\mathscr{E}\right](t,T+\tau) =
    D^{(1)}_{\mu\nu}
    \left[\mathscr{E}, \delta_{T+\tau}\right](t) \,.
    \label{eq:first_order_response_deltafunction}
\end{equation}
This object contains all the necessary information about the interacting system to compute the absorption
of any given probe, as long as it is weak enough for the response to be linear. In 
order to analyze it, it is useful to take the Fourier transform with respect
to the variable $t$, and expand
Eq.~\eqref{eq:retarded_response_function_recast} in an eigenbasis
of the static Hamiltonian $\mathscr{\hat H}$ 
\begin{equation}
    \label{eq:wavefunction_eigenstates}
   \vert\Psi\left[\mathscr{E}\right](T)\rangle = \SumInt_{j=1}^{\infty}\gamma_{j} |\Phi_j\rangle , \qquad \mathscr{\hat
    H}|\Phi_j\rangle=\varepsilon_j |\Phi_j\rangle \,,
\end{equation}
obtaining a Lehmann representation for the time-dependent non-equilibrium response function:
\begin{eqnarray}\label{eq:generalized_lehmann_representation_real_operators}
    \chi_{\hat{D}_\mu, \hat{D}_\nu}&&\left[\mathscr{E}\right](\omega, T+\tau) =\\ 
    &&\SumInt_{jkm} d^\mu_{jm}d^\nu_{mk} 
       \left\{ 
         \frac{\gamma_{j}\gamma_{k}^*e^{i\omega_{jk}\tau}}{\omega+\omega_{jm}+i\sfrac{\Gamma}{2}}
        -\frac{\gamma_{j}^*\gamma_{k}e^{-i\omega_{jk}\tau}}{\omega-\omega_{jm}+i\sfrac{\Gamma}{2}}
            \right\} \nonumber
\end{eqnarray}
in terms of the exact energy differences $\omega_{jk}=\varepsilon_k-\varepsilon_j$, 
the dipole-matrix elements $d^\mu_{jm}=\langle\Phi_j|\hat D_\mu|\Phi_m\rangle\in\mathbb{R}$
and the pump-probe delay $\tau$.
A similar representation in the case of non-equilibrium
spectroscopy %
has been recently presented in a similar context~\cite{perfetto-2015}.
By writing $\gamma_j=|\gamma_j|e^{i\varphi_j}$ Eq.~\eqref{eq:generalized_lehmann_representation_real_operators} turns into
\begin{align}
    \label{eq:lehmann2}
    &\chi_{\hat D_\mu, \hat D_\nu}\left[\mathscr{E}\right](\omega, T+\tau) =  \\
    &\SumInt_{jkm} d^\mu_{jm}d^\nu_{mk} |\gamma_j\gamma_k|
        \left\{ 
        \frac{e^{i\Theta_{kj}(\tau)}}{\omega+\omega_{jm}+i\sfrac{\Gamma}{2}}
        -\frac{e^{-i\Theta_{kj}(\tau)}}{\omega-\omega_{jm}+i\sfrac{\Gamma}{2}}
            \right\}\nonumber
\end{align}
with
\begin{equation}
  \Theta_{kj}(\tau)=\varphi_j-\varphi_k-\omega_{kj}\tau \, .
  \label{eq:theta}
\end{equation}
Since the absorption
depends on the imaginary part of the response function, we get from Eq.~\eqref{eq:lehmann2}:
\begin{align}\nonumber
\Im    \chi_{\hat D_\mu, \hat D_\nu}\left[\mathscr{E}\right](\omega, T+\tau) =  
\SumInt_{jkm} d^\mu_{jm}d^\nu_{mk} |\gamma_j\gamma_k|
\\\nonumber
\left \{ \cos (\Theta_{kj}(\tau))L(\omega-\omega_{jm}) + \sin(\Theta_{kj}(\tau))R(\omega-\omega_{jm}) \right. \\\nonumber
\left.  - \cos (\Theta_{kj}(\tau))L(\omega+\omega_{jm}) + \sin(\Theta_{kj}(\tau))R(\omega+\omega_{jm}) \right\} \\
\label{eq:chi_final}
\end{align}
with the Rayleigh peaks
\begin{equation}
    R(\bar\omega)=\frac{\bar \omega}{\bar\omega^2+\sfrac{\Gamma^2}{4}}
    \label{eq:rayleigh}
\end{equation}
and the Lorentzian peaks 
\begin{equation}
    L(\bar\omega)=\frac{\sfrac{\Gamma}{2}}{\bar\omega^2+\sfrac{\Gamma^2}{4}} \, .
    \label{eq:lorentzian}
\end{equation}
The absorption cross section tensor is:
\begin{equation}
\sigma_{\mu, \nu}(\omega)=\frac{4\pi\omega}{c}\Im\chi_{\hat D_{\mu}\hat D_{\nu}}(\omega)\,,
\end{equation}
and for a random sample, the absorption will be its orientational average,
i.e.\ the \emph{absorption coefficient}:
\begin{equation}
\label{eq:absorption-coefficient}
  \bar\sigma(\omega)=\frac{1}{3}\Tr\sigma(\omega)\,.
\end{equation}
Since we are only interested in the trace, in the following, we concentrate on the diagonal terms,
we will omit hereafter the orientation indexes in order to ease the notation. 

We now analyze Eq.~\eqref{eq:chi_final} in more detail. First,
as already pointed out in
Ref.~\cite{perfetto-2015}, in this non-overlapping
regime the dependence of the spectrum on both the pump-pulse and the delay time enters through the modification of the
peak amplitudes and shapes exclusively: the peak positions are intrinsic properties of the many-body system.
Second, for the analysis of the effect of the pump pulse and of the pump-probe delay
on the spectrum, we can distinguish between three cases: 
(i) the system is in its ground state,  
(ii) the system is in an excited eigenstate,
(iii) the system is in a non-stationary state, i.e.\ in a linear combination of non-degenerate eigenstates.
In all cases we focus on the positive part of the energy range, which we
denote by $\omega^+$. The following shape analysis in 
terms of Lorentzian and Rayleigh contributions is for discrete peaks only. We denote this by replacing $\SumInt$ by $\sum$
(we will comment on the continuum part later on in this section).

In case (i) $\gamma_i=\delta_{i0}$ and $\Im \chi_{\hat D, \hat D}\left[\mathscr{E}\right](\omega^+)$
reduces to the usual Lehmann representation for the ground state spectrum:
\begin{equation}
\Im \chi_{\hat D, \hat D}\left[\mathscr{E}\right](\omega^+) =  \sum_{m} |d_{0m}|^2 L(\omega^+-\omega_{0m}).
\label{eq:chi_case1}
\end{equation}
All peaks are positive and have Lorentzian shape. 
In case (ii), where $\gamma_i=\delta_{i \xi}$, the system is in an excited
eigenstate  $\Phi_\xi$, 
and therefore in the positive energy part of the spectrum we can find both positive and negative peaks of Lorentzian shape:
\begin{eqnarray}\label{eq:chi_case2}
 \Im    \chi_{\hat D, \hat D}&&\left[\mathscr{E}\right](\omega^+) =\\  
&&  \sum_{\xi < m} |d_{\xi m}|^2 L(\omega^+-\omega_{\xi m})  
- \sum_{\xi > m} |d_{\xi m}|^2 L(\omega^++\omega_{\xi m}) \, .\nonumber
\end{eqnarray}
Note, that in both cases (i) and (ii), the spectrum is time-independent and has only Lorentzian contributions. This is
the main difference to the case (iii), for non-stationary states.
In this case, the spectrum can be divided into two parts, 
one time-independent and one oscillatory part due to interferences between the
involved states:
\begin{eqnarray} \label{eq:im_partitioning}
    \Im \chi_{\hat D, \hat D}&&\left[\mathscr{E}\right](\omega^+, T+\tau) = \\
&&        \Im \chi_{\hat D, \hat D}^{0}\left[\mathscr{E}\right](\omega^+) 
    +   \Im \chi_{\hat D, \hat D}^{\rm INT}\left[\mathscr{E}\right](\omega^+, T+\tau)   \nonumber
\end{eqnarray}
The equilibrium term consists of the sum over the stationary state spectra of the eigenstates involved, scaled by their
occupations:
\begin{eqnarray}
  \Im &&   \chi_{\hat D, \hat D}^{0}\left[\mathscr{E}\right](\omega^+) =
  \sum_j |\gamma_j|^2  \times \\
  && \left\{
  \sum_{j < m} |d_{j m}|^2 L(\omega-\omega_{j m})  
  - \sum_{j > m} |d_{j m}|^2 L(\omega+\omega_{j m}) \right\} \, .\nonumber
    \label{eq:chi_equi1}
\end{eqnarray}
Its peaks are always of Lorentzian shape and depend neither on time nor on the initial phase difference
$\varphi_j-\varphi_k$. It is influenced by the pump laser only through the occupations $|\gamma_j|^2$. 
The phase- and time-dependency of the spectrum enters through the interference term
\begin{eqnarray}\nonumber
 &&\Im  \chi_{\hat D, \hat D}^{\rm  INT}\left[\mathscr{E}\right](\omega, T+\tau) =  
\sum_{j\neq k; m}  d_{jm}d_{mk}  |\gamma_j\gamma_k| \left \{ \right. \\\nonumber
&& \cos \Theta_{kj}(\tau)L(\omega-\omega_{jm}) + \sin\Theta_{kj}(\tau)R(\omega-\omega_{jm})  \\\nonumber
&-&\left. \cos \Theta_{kj}(\tau)L(\omega+\omega_{jm}) + \sin\Theta_{kj}(\tau)R(\omega+\omega_{jm}) \right\} \\
    \label{eq:chi_nonequi}
\end{eqnarray}
which in turn is governed by the phase differences $\Theta_{kj}$, which have contributions from both the
phase difference $\varphi_j-\varphi_k$ at the end of the laser and from its time evolution $\omega_{kj}\tau$.
$\Theta_{kj}$ mixes real and imaginary part of the response function and converts Lorentzian line shapes into Rayleigh line shapes and 
vice versa. This conversion happens periodically with the frequency given by the energy differences $\omega_{kj}$
between the occupied states involved. 
The time-dependence of a spectrum is therefore a clear sign of a non-stationary
state. Experimentally, this periodic beating pattern was recently observed by Goulielmakis \emph{et al}
\cite{goulielmakis-2010}.
This demonstrates, how using a pump to imprint an internal phase difference $\varphi_j-\varphi_k$ onto a state and
controlling the delay time $\tau$ between pump and probe laser can be used to change a spectrum, converting
absorption into emission peaks (and vice versa) as well as changing the overall shape of the lines.
In Section~\ref{sec:hydrogen} we demonstrate these line shape changes using the example of an exactly solvable Hydrogen
atom. Furthermore we demonstrate, how to use a laser to control these features.

Note, that in the discussion above, the lineshape analysis is valid for isolated peaks without contribution from
continuum states. If coupling to continuum states is involved, an additional shaping comes from the
dependence of the matrix elements $d_{jm}$ on the energy. This is e.g.\ the case for Fano line-shapes which 
may acquire a complex Fano $q$ factor~\cite{Zielinski:2014vx}. 

\section{Quantum Optimal Control of Excited State Spectra}\label{sec:qoct}

In this work we employ Quantum Optimal Control Theory (QOCT) to optimize the
response of a system in the situation described in the previous section.
QOCT is concerned with studying the optimal Hamiltonian (in practice, a
portion of the Hamiltonian, such as the temporal profile of the coupling of an
atom or molecule to a laser pulse) that induces a target system behaviour.
In the following, we present its specific application to the problem of optimizing
response functions of excited states. 
We will also show how, if the problem can be reduced to a small model, it can be solved 
analytically.

Let us consider a quantum mechanical system governed by the
Schr{\"{o}}dinger equation during the time interval [0,T]:
\begin{subequations}
    \label{eq:schroedinger_control}
\begin{eqnarray}
\label{eq:schroedinger-1}
i\frac{\partial \Psi}{\partial t} (x,t)  & = & \hat{H}[\mathscr{E},t] \Psi(x,t) \,,
\\
\label{eq:schroedinger-2}
\Psi(x,0) & = & \Psi_0(x)\,,
\end{eqnarray}
\end{subequations}
where $x$ is the full set of quantum coordinates, and $\mathscr{E}$ is the
\emph{control} field, an external potential applied to the
system (in our case, the pump pulse). 
In order to perform optimizations the field must be discretized, 
for example with the help of a sine Fourier basis. In our numerical simulations:
\begin{equation}
    \mathscr{E}_{\mathbf{c}}(t) = \sum_{n=1}^M  c_n \sin(\omega_n t)
    \label{eq:controlfield}
\end{equation}
where $M$ is the dimension of the optimization search space, and $\mathbf{c}$ is
the set of all the parameters that determine the field: $\mathbf{c} = c_1,\dots,c_M$. The frequencies,
and their maximum value or cut-off frequency, may be chosen at will. 

The specification of $\mathscr{E}$, together with an initial value
condition, $\Psi(0)=\Psi_0$ determines the full evolution of the system,
$\Psi[\mathscr{E}]$, via the propagation of the Schr{\"{o}}dinger equation.
The behaviour of the system must then be measured by defining a ``target
functional'' $F$, whose value is high if the system evolves according to our
goal, and small otherwise. In many cases, it is split into two parts,
$F[\Psi,\mathscr{E}] = J_1[\Psi] + J_2[\mathscr{E}]$, so that $J_1$ only depends on the state of
the system, and $J_2$, called the ``penalty'', depends explicitly on the
control $\mathscr{E}$. Regarding $J_1$, it may depend 
on the full evolution of the system during the time interval $[0,T]$, or only
on the system state at time $T$, as it is the case in this work.
Often, the functional is defined through the expectation value
of an observable $\hat{O}$:
\begin{equation}
J_1^{T}[\Psi(T)] = \langle \Psi(T)\vert\hat{O}\vert\Psi(T)\rangle\,.
\end{equation}

The mathematical problem is then reduced to the problem of maximizing a
real-valued function $G$:
\begin{equation}
G[\mathbf{c}] = F[\Psi[\mathscr{E}_{\mathbf{c}}],\mathscr{E}_{\mathbf{c}}]\,.
\end{equation}
The absorption of light is related to the average absorption coefficient 
$\bar\sigma[\mathscr{E}\mathbf{c}](E)$ [Eq.~(\ref{eq:absorption-coefficient})]. The larger the absorption coefficient
at a certain energy, the more light is absorbed at this energy. 
In order to find a laser pulse to make a system, that is 
transparent in its ground state, absorb as much light as possible, we therefore optimize the absorption coefficient 
in the visible by taking the integral of $\bar\sigma[\mathscr{E}_\mathbf{c}](E)$ 
over the respective energy range. We employed two different control targets:
\begin{subequations}
\begin{eqnarray}
    G_{\tau}^{A}\left[\mathbf{c}\right]&=&\int_{E_{min}}^{E_{max}}\!\!\!\!  dE \, \,  \bar\sigma_{\tau}[\mathscr{E}_\mathbf{c}](E)\,,\\
    \label{eq:epsilon_a}
    G_{\tau}^{B}\left[\mathbf{c}\right]&=&\int_{E_{min}}^{E_{max}}\!\!\!\! dE \, \, \bar\sigma_{\tau}[\mathscr{E}_\mathbf{c}](E)
    e^{\left(-\gamma\frac{N_0-N_{T}[\mathscr{E}]}{N_0}\right)} \, ,
    \label{eq:epsilon_b}
\end{eqnarray}
    \label{eq:def_epsilon}
\end{subequations}
where $\bar\sigma_{\tau}[\mathscr{E}_\mathbf{c}](E)$  [in the following we will call it just $\bar\sigma(E)$]
is the average absorption coefficient 
of the system at a given time delay $\tau$ after the pump
pulse $\mathscr{E}(t)$, and $E_{min}$ and $E_{max}$ define the optimization
region - the energy range, where the absorption is optimized. In the second
target function we have introduced an exponential factor that depends on
$N_0$ and $N_{T}$, the number of electrons in the system at the
beginning and the end of the pump pulse, respectively. The reason to
introduce this factor is to avoid ionization, i.e.\ we wish to lead the system
to a state with the desired absorption properties, but keeping the ionization
probability low. Keeping the ionization low is particularly important in the TDDFT calculations, if performed with
adiabatic functionals, since with current state-of-the-art adiabatic functionals, 
ionization of the system leads to unphysical
shifts in the position of the absorption peaks. The term $\exp
{\left(-\gamma\frac{N_0-N_{T}[\mathscr{E}]}{N_0}\right)}$ therefore inflicts
a penalty, whose strength can be modulated by $\gamma$, to pump pulses that
produce strong ionization. We implement ionization using absorbing boundaries and thus the
total number of electrons is, in general, not conserved during time.
In practice, one can also combine the two target functions: one may start
optimizations using $ G_{\tau}^{A}$,
and later continue with $G_{\tau}^{B}$,
restarting from the previous optimum.

Once the target is defined one is left with the problem of 
choosing an optimization algorithm to find the maximum (or
maxima) of $G$. Two broad families can be distinguished: gradient-free
procedures, which only require the computation of $G$
given a control input $\mathscr{E}$, and gradient-based procedures, that also
require the computation of the gradient of $G$ with respect to $\mathscr{E}$.
QOCT provides an expression for the gradient that can be adapted for this
case (see appendix \ref{sec:qoct_gradientbased} for details).
This approach, however is numerically unfeasible for the target covered in this paper.
For this reason, in our simulations, we employed the gradient-free Simplex-Downhill algorithm by Nelder 
and Mead~\cite{nelder-1965}.

In principle, if the system can be reduced to a few-level model, the optimal
fields can be found analytically. To illustrate this approach, below we briefly illustrate a 
simple example of controlling the absorption properties of a single Hydrogen atom. 
Instead of directly optimizing $G_{\tau}^{A,B}$
we here derive a laser that drives the system into a state with the wanted optical properties.
Let us suppose that the situation can be approximated by a three-level Hamiltonian $\mathscr{\hat H}$
with eigenstates $|\Phi_a\rangle$, $|\Phi_b\rangle$ and $|\Phi_c\rangle$ and the corresponding eigenenergies $\varepsilon_a$,
$\varepsilon_b$ and $\varepsilon_c$. 
We define the transition energies $\omega_{ab}=\varepsilon_b-\varepsilon_a$, $\omega_{bc}=\varepsilon_c-\varepsilon_b$ and
$\omega_{ac}=\varepsilon_c-\varepsilon_a$.
The dipole coupling between the states is given by $d_{ab}$ and $d_{ac}$ (both
assumed to be real numbers), and we further consider the case where the coupling between the states 
$|\Phi_b\rangle$ and $|\Phi_c\rangle$ is dipole forbidden.

The system is pumped by a laser field composed of two carrier frequencies $\omega_{1,2}$ of the form:
\begin{equation}
    \mathscr{E}(t)=\tilde{\varepsilon}_1(t)\cos(\omega_1 t+\varphi_1) + \tilde{\varepsilon}_2(t)\cos(\omega_2 t + \varphi_2),
\end{equation}
with phases $\varphi_{1,2}$, amplitudes $\varepsilon_{1,2}$ and envelope $\tilde{\varepsilon}_{1,2}(t)$ defined by
\begin{eqnarray}
    \tilde\varepsilon_{1,2}(t) &=& 
                    2\varepsilon_{1,2} \sin^2\left(\pi\frac{t}{T}\right)\,.
\end{eqnarray}
Our goal is to find a laser pulse that drives the system from state
$|\Psi(t=0)\rangle=|\Phi_a\rangle$ into a target state
$|\bar{\Psi}\rangle$
\begin{equation}
    |\bar{\Psi}\rangle=\alpha|\Phi_a\rangle+\beta|\Phi_b\rangle+\gamma|\Phi_c\rangle
 \label{eq:hcontrol_target}
\end{equation}
in a given time $T$ -- $\alpha,\beta,$ and $\gamma$ are complex coefficients. 
Since the spectral properties of this state can be then easily obtained 
using Eqs.~\eqref{eq:im_partitioning}, \eqref{eq:absorption-coefficient},   
the problem of finding a pulse giving the desired optical properties translates to 
the one of maximizing the overlap 
$|\langle\Psi(T)|\bar{\Psi}\rangle|^2$ while keeping the functional form of the laser fixed -- i.e.\ changing only 
$\omega_{1,2}$, $\varphi_{1,2}$, and $\varepsilon_{1,2}$.
If we choose $\omega_{1,2}$ resonant with the transition frequencies $\omega_{ab},\omega_{ac}$ and
assume they are sufficiently separated in energy we can apply the rotating wave
approximation and obtain the laser parameters as function of $\alpha,\beta,\gamma$
as (see Appendix~\ref{sec:analyticalcontrol} for details):
\begin{subequations}
\begin{eqnarray}
    \varepsilon_1 &=& \frac{2}{T}\frac{\arccos(|\alpha|)}{\sin(\arccos(|\alpha|)}\frac{|\beta|}{d_{ab}}
\\
    \varepsilon_2 &=& \frac{2}{T}\frac{\arccos(|\alpha|)}{\sin(\arccos(|\alpha|)}\frac{|\gamma|}{d_{ac}}
 \end{eqnarray}
and
\begin{eqnarray}
    \varphi_{\beta} &=&  \varphi_1-\pi+\omega_{ba}T \\
    \varphi_{\gamma} &=& \varphi_2-\pi+\omega_{ca}T \,.
\end{eqnarray}
\end{subequations}
We will come back to this example below in Sec.~\ref{sec:hydrogen}.

\section{Applications}\label{sec:applications}

Any QOCT formulation is constructed on top of a given model for the physics of the
process under study. In this paper we study and optimize the absorption spectra of atoms and molecules
using either analytically solvable model Hamiltonians or obtaining the spectra by using time-dependent density
functional theory (TDDFT)~\cite{ book-tddft,tddft-specialissue} -- the time-dependent counterpart of DFT~\cite{kohn-1999}. 

Based on the Runge-Gross theorem~\cite{runge-1984} TDDFT establishes a one-to-one correspondence between
the time-dependent density and the time-dependent external potential of a many-electron system.
Together with the Kohn-Sham (KS) scheme \cite{kohn-1965} it allows us to recast the many-body time-dependent 
problem into a simpler one where the interacting electrons are replaced by a 
fictitious set of non-interacting electrons with the same time-dependent density.
This system of non-interacting electrons can then be 
represented with a single Slater determinant formed by a set of KS orbitals 
leading to great computational simplifications.

In the following we will work with spin-compensated systems of $N$ electrons doubly occupying $N/2$ spatial orbitals.
The time evolution of these orbitals $\varphi_i$ ($i=1,N/2$), is governed by the time-dependent Kohn-Sham equations 
\begin{eqnarray}
{\rm i}\frac{\partial}{\partial t}\varphi_i(\vec{r},t) & = & -\frac{1}{2}\nabla^2 \varphi_i(\vec{r},t)
+ v_{\rm KS}[n](\vec{r},t)\varphi_i(\vec{r},t)\,,
\\
n(\vec{r},t) & = & 2\sum_{i=1}^{N/2}\vert\varphi_i(\vec{r},t)\vert^2\,,
\end{eqnarray}
\label{eq:tdks}
where $v_{\rm KS}[n](\vec{r},t)$ is the KS potential. 
It is, in general, a functional of the density and is defined as 
\begin{equation}
v_{\rm KS}[n](\vec{r},t) = v_0(\vec{r}) + v(\vec{r},t) + v_{\rm H}[n](\vec{r},t) + v_{\rm xc}[n](\vec{r},t)\,,
\end{equation}
where $v_0(\vec{r})$ represents the static (ionic) external potential, $v(\vec{r},t)=\mathcal{E}(t)\cdot {\vec r}$ 
is the coupling to the time dependent electric field $\mathcal{E}(t)$ in the dipole approximation (in the length gauge),
$v_{\rm H}[n](\vec{r},t) = \int\!{\rm d}^3r'\,n(\vec{r},t)/\vert \vec{r}'-\vec{r}\vert$ is the classical electrostatic Hartree potential,
and $v_{\rm xc}[n](\vec{r},t)$ is the exchange and correlation potential accounting for the many electron effects~\cite{book-tddft,marques-2012}.
In our simulations the ions are clamped to their equilibrium positions.
All numerical calculations were performed using the octopus code
\cite{marques-2003}.

\subsection{One Electron Systems: the Hydrogen Atom}\label{sec:hydrogen}
In Section~\ref{sec:spec_ex} we have discussed how the amplitudes and shapes of
excited state absorption spectra
depend on the relative phases $\varphi_i$ of the expansion coefficients
$\gamma_i$. Here, we illustrate this effect
in a Hydrogen atom, which is initially pumped into the state  
\begin{equation}
    |\bar{\Psi}\rangle=\sqrt{0.4}|2p_z\rangle + \sqrt{0.6}e^{i\varphi}|3p_z\rangle\,.
    \label{eq:hydrogen_target}
\end{equation}
Let us first vary $\varphi$ to show the effect of the phase on the final spectrum. 
For our pumped state the stationary part of the spectrum 
is composed of the weighted stationary state spectra coming from $|2p_z\rangle$ and $|3p_z\rangle$:
\begin{equation}
  \bar\sigma^0(\omega)=0.4\bar\sigma_{2p_z}(\omega)+0.6\bar\sigma_{3p_z}(\omega)\,,
\end{equation}
whereas the phase-dependent interference term is
\begin{align}\nonumber
&  \bar\sigma^{\rm INT}(\omega, \varphi) =  0.4\cdot 0.6\cdot
  \frac{4\pi\omega}{3c}\; \Big[
\SumInt_{m} d_{2p,m}d_{m, 3p}
\\\nonumber
&           \left \{ \cos \varphi L(\omega-\omega_{3p,m}) + \sin\varphi
R(\omega-\omega_{3p,m}) \right. 
\\\nonumber
&           \left.  - \cos\varphi L(\omega-\omega_{3p,m}) + \sin\varphi
R(\omega-\omega_{3p,m}) \right\} 
\\\nonumber
& +\SumInt_{m} d_{2p,m}d_{m, 3p}
\\\nonumber 
&           \left\{ \cos \varphi L(\omega-\omega_{2p,m}) - \sin \varphi
R(\omega-\omega_{2p,m}) \right. 
\\
&          \left. -\cos \varphi L(\omega-\omega_{2p,m}) - \sin \varphi
R(\omega-\omega_{2p,m}) \right\} \Big]
    \label{eq:chi_nonequi_h}
\end{align}
Note the change in the sign of the Rayleigh terms in both sums.
\begin{figure}[htb!]
    \begin{center}
        \includegraphics[clip, width=\columnwidth]{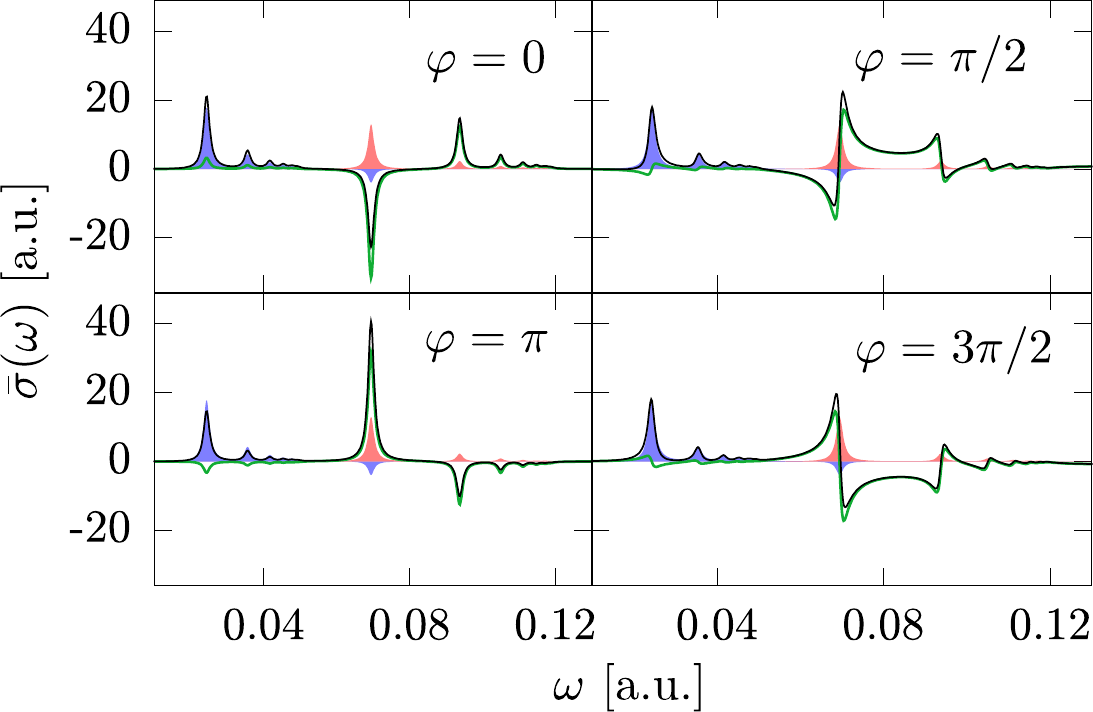}
    \end{center}
    \caption{Absorption coefficient $\bar\sigma(\omega)$ of the state  
    defined in Eq.~\eqref{eq:hydrogen_target} with $\varphi=0$, $(1/2)\pi$, $\pi$ and $(3/2)\pi$.
    The total spectrum (black line) is the sum of two phase-independent terms $0.4\bar\sigma_{2p_z}$ (red shaded) 
    and $0.6\bar\sigma_{3p_z}$ (blue
    shaded) coming from the excited state spectra of the respective states,
    plus the phase-dependent interference term 
    $\bar\sigma^{\mathrm{INT}}(\omega, \varphi)$ (green dashed line), which is responsible for the change of the
    spectrum with the delay time.}
    \label{fig:H_spectrum_analytic_phases}
\end{figure}
Fig.~\ref{fig:H_spectrum_analytic_phases} shows the different contributions and the complete spectrum for 
$\varphi = 0$, $(1/2)\pi$, $\pi$ and $(3/2)\pi$, which are the cases, where the interference term is either
purely Lorentzian ($\varphi=0$, $\pi$) or purely Rayleigh ($\varphi=1/2\pi$, $3/2\pi$). 
The shaded areas
indicate the weighted equilibrium contributions, the dotted line
shows the interference terms, and the solid line the complete spectrum. 
The energy range shown includes the transitions from
n = 2 to all higher states and from n = 3 to all higher states and to n = 2. 
Transitions to the ground state lie outside of this region. 

As can be learnt from Eq.~\eqref{eq:chi_nonequi_h}, the interference terms require the existence of states
which are dipole-coupled to both $2p_z$ and $3p_z$. 
This is the case for $s$- and $d$-orbitals. This means, that e.g.\ for Hydrogen in a linear
combination of the states $2s$ and $4f$, all the
interference terms would vanish and the spectrum would be purely the sum of the weighted equilibrium contributions.

Let us take a closer look at the structure of the interference terms. 
We start with the interference term at $\omega_{23}=0.069$ Ha having contributions from terms with $m=2s$, $m=3s$ and $m=3d$. 
All contributions have different prefactors with the ones coming from the 2s-state 
having the opposite sign compared to the ones coming from 3s and 3d states. 
For the other peaks, the interference terms are much smaller at the 
energies $\omega_{3n}$ than their counterparts at 
$\omega_{2n}$ (compare the purely blue to the purely red peaks in Fig.~\ref{fig:H_spectrum_analytic_phases}).
From Eq.~\eqref{eq:chi_nonequi_h}, it is apparent that the amplitude of each interference term
is the same for $\omega_{2n}$ and $\omega_{3n}$ with the same $n$. The difference comes purely
from the factor $\frac{4\pi\omega}{3c}$ -- note that the sign of the Rayleigh
contributions is opposite in these pairs of peaks. 
This variation of amplitude has the following consequences for the change of 
the overall spectrum:
At $\omega_{3n}$ the spectrum has
positive contributions from $\bar\sigma_{3p_z}$ and contributions from the interference terms, 
but since the interference terms are much
smaller than $\bar\sigma_{3p_z}$, the spectrum changes only slightly for different $\varphi$'s. 
This is different for the peaks at energies $\omega_{2n}$. 
Here, the spectrum has positive, phase-independent contributions from
$\bar\sigma_{2p_z}(\omega)$, but the contributions from the interference terms are much larger 
and dominate the spectrum leading
to a strong dependence of the spectrum in this energy range on the phase $\varphi$.
For $\varphi=0$ and $\varphi=\pi$, 
$\bar\sigma^{\rm INT}(\omega, \varphi)$ only contains Lorentzian peaks and consequently 
the whole spectrum only contains Lorentzians. Nevertheless, $\bar\sigma^{\rm INT}(\omega, \varphi)$ 
changes sign between $\varphi=0$ and $\varphi=\pi$, switching the sign of all
peaks at $\omega_{2n}$. This is a
demonstration of, how the manipulation of the internal phase $\varphi$ can lead to a switch from gain (negative peaks)
to loss (positive peaks) regime and
vice versa. Finally, for $\varphi=(1/2)\pi$ and $\varphi=(3/2)\pi$, the interference spectrum contains purely Rayleigh
peaks. Together with the small contributions from the stationary-state contributions, the final spectrum consists
of slightly asymmetric Rayleigh peaks, again with different signs for $\varphi=(1/2)\pi$ and $\varphi=(3/2)\pi$. One can
therefore not only change peaks from emission to absorption peaks, but also manipulate their shape.
The phases $\varphi$ therefore play a critical role in the spectral weights and the
peaks of the photo-absorption spectrum. 

\begin{figure}[htb!]
     \begin{center}
       \includegraphics[width=\columnwidth]{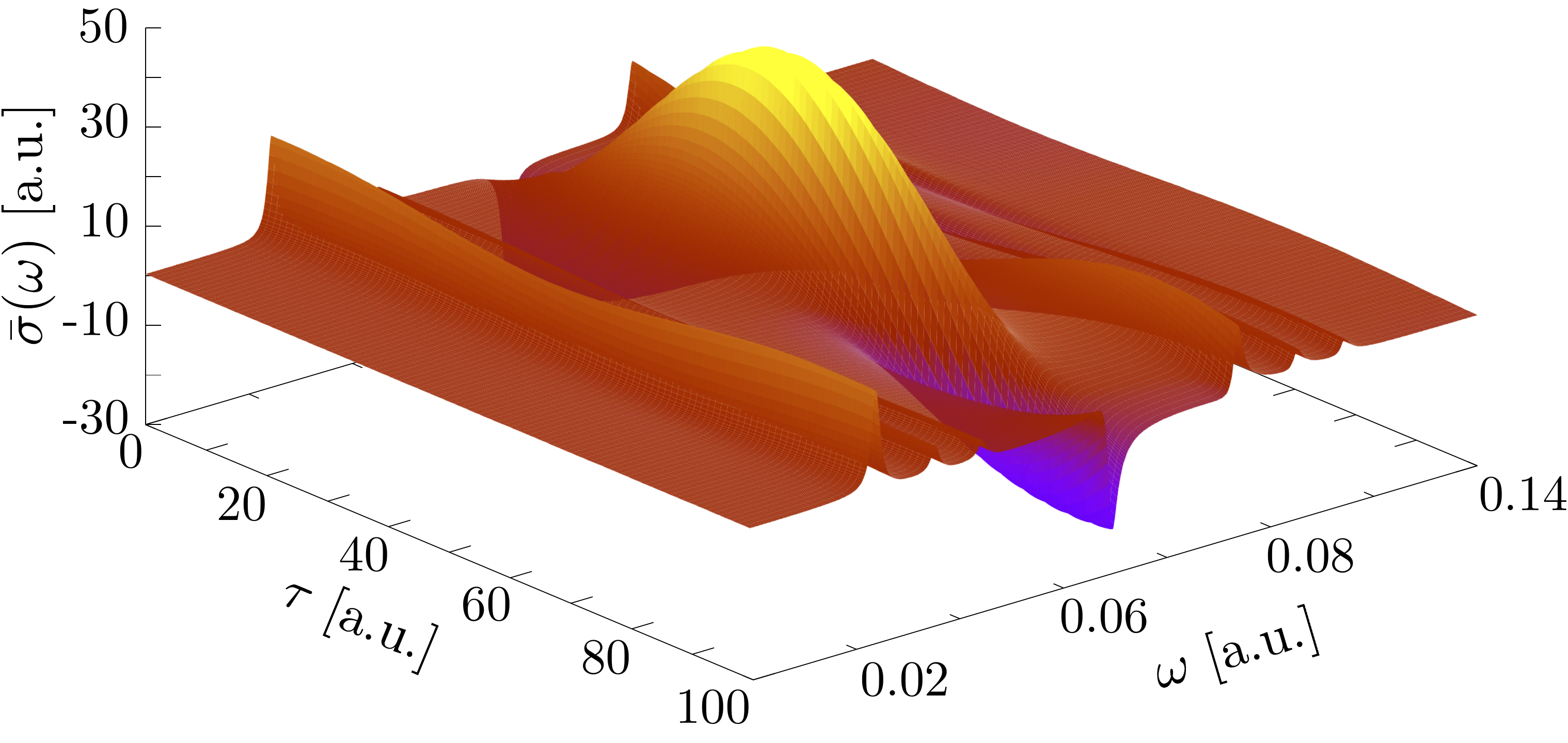}
     \end{center}
    \caption{Time-resolved spectrum of the initial state 
    $\sqrt{0.4} |2p_z\rangle + \sqrt{0.6} |3p_z\rangle$ of Hydrogen. Because the
    phases of the $|2p\rangle$ and $|3p\rangle$ states evolve with different velocities, 
    the spectral weights of each of the peaks changes with time, leading to a time-dependent 
    spectrum with a periodicity of $T = \frac{2\pi}{\omega_{32}} \approx 91 a.u.$. 
    }
    \label{fig:Hydrogen_spectroscopy_td_spectrum}
\end{figure}

We now look at the variation of the spectrum with time, 
assuming that an initial, yet unknown pump laser created the state of Eq.~\eqref{eq:hydrogen_target} with $\varphi=\varphi_{32}=0$ at $t=T$
and we probe the system at different delay times $\tau$.
Figure~\ref{fig:Hydrogen_spectroscopy_td_spectrum} shows the corresponding time-resolved spectrum $\bar\sigma(\omega, \tau)$ 
of $|\bar{\Psi}\rangle$. Since the eigenenergies of $|2p_z\rangle$ and $|3p_z\rangle$ are different, 
the phase $\Theta_{32}(\tau)$ in Eq.~\eqref{eq:theta}
evolves with the frequency $\omega_{32}$. At $\tau=0$, $\tau=\frac{\pi}{2\omega_{32}}$, $\tau=\frac{\pi}{\omega_{32}}$
and $\tau=\frac{3\pi}{2\omega_{32}}$, the spectra of Fig.~\ref{fig:H_spectrum_analytic_phases} are reproduced. 
One sees the strong changes of $\bar\sigma$ in the energy range of the peaks $\omega_{2n}$, 
while the peaks $\omega_{3n}$ remain almost unchanged. 
The spectrum is periodic with $T=\frac{2\pi}{\omega_{32}}\approx$ 91~a.u..

We now move on to the control problem -- i.e.\ the design of a pump pulse
driving the system into a state with specific optical properties.
For this problem, we will use the three-levels model, and the analytical
equations of control presented in Section~\ref{sec:qoct}.
The target state will again be the one defined in Eq.~\eqref{eq:hydrogen_target} 
$|\bar{\Psi}\rangle=\sqrt{0.4}|2p_z\rangle + \sqrt{0.6}e^{i\varphi}|3p_z\rangle$
with a relative phase of $\varphi=0$: $|\bar{\Psi}\rangle=\sqrt{0.4}|2p_z\rangle + \sqrt{0.6}|3p_z\rangle$.
The three active states are then 
$|1s\rangle$, $|2p_z\rangle$ and $|3p_z\rangle$. Note that since
$|2p_z\rangle$ and $|3p_z\rangle$ have the same
symmetry, they are decoupled in the dipole approximation, and in consequence the system
fits into the framework described in Section~\ref{sec:qoct}. 
We may therefore write down the shape of a control pulse,
assuming a total pulse time of $T=3200$~a.u:
\begin{eqnarray}    \label{eq:hcontrol_laser_numerical}
    \mathscr{E}(t) = \frac{2\pi}{T}  
       && \left(  \frac{\sqrt{0.4}}{d_{1s\rightarrow 2p}}\cos(\omega_{1s\rightarrow 2p}(t-T)+\pi) \right.  \\
      +&& \left.  \frac{\sqrt{0.6}}{d_{1s\rightarrow 3p}}\cos(\omega_{1s\rightarrow 3p}(t-T)+ \pi)\right)
                    \sin^2\left(\pi\frac{t}{T}\right).\nonumber
\end{eqnarray}
We numerically solved the TSDE in order to check the validity of the three-level approximation. 
To this end we discretized the equations on a spherical grid of radius $R=60$~a.u., 
spacing of $\Delta x = 0.435$~a.u.\ and with 20~a.u.\ wide absorbing boundaries placed at the edges.
\begin{figure}[htb!]
    \begin{center}
        \includegraphics[width=\columnwidth]{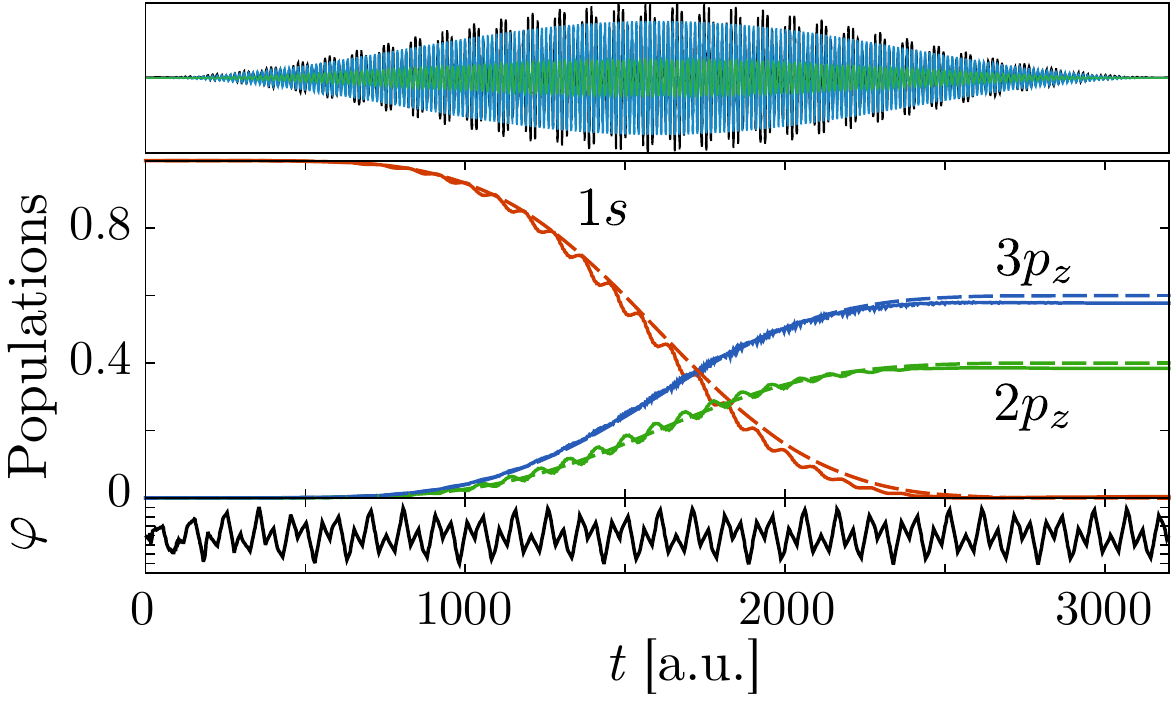}
    \end{center}
    \caption{Time-evolution of the populations of the $1s-$, $2p_z-$ and $3p_z-$ state. Dashed lines show the 
    analytic model, solid lines the numerical results. The total pump-laser (upper panel, black) has two carrier-frequencies, 
    one resonant to the transition $|\Psi_{1s}\rangle\rightarrow|\Psi_{2p}\rangle$ (green), the other resonant to the
    transition $|\Psi_{1s}\rangle\rightarrow|\Psi_{3p}\rangle$ (blue). The lower panel shows the phase difference
    $\varphi_{3p}-\varphi_{2p}$.}
    \label{fig:H_spectra_projections}
\end{figure}
The results are collected in Figure~\ref{fig:H_spectra_projections} where we show the time-evolution of the 
populations $|a(t)|^2$, $|b(t)|^2$ and $|c(t)|^2$ of the states 
$|1s\rangle$, $|2p_z\rangle$ and $|3p_z\rangle$ respectively. The numerical values (solid lines)
follow closely the ones corresponding to the model
\eqref{eq:hcontrol_populations_td} (dashed lines) except for a small superimposed oscillatory behavior.
A frequency analysis of the additional 
oscillations shows, that they are due to the components neglected in the rotating wave approximation. The
small deviation in the final populations from the analytic prediction comes from the excitation into the
3d-states (not shown). The coupling to these orbitals was neglected in the three-level approximation.
This population transfer to the 3d-states nonetheless is less than 4\% , and
we achieve a transfer into
the target wave function $|\bar{\Psi}\rangle$ of 96\%. 
Furthermore the transfer is obtained precisely with the desired relative phase $\varphi=0$ as reported in the 
bottom panel of Figure~\ref{fig:H_spectra_projections}.

\subsection{More Than one Electron: Results Based on TDDFT}\label{sec:tddft}
We here turn to systems with more than one electron, and investigate
the possibility to drive the absorption of atoms and molecules into the visible using a 
laser pulse optimized with the gradient-free optimization algorithm presented in Section~\ref{sec:qoct}
in combination with TDDFT.

\subsubsection{Helium}
As a first example we study the one-dimensional soft-Coulomb Helium atom.
This model is defined by the Hamiltonian:
\begin{equation}
H(t)= T + V_{\rm ext}(t)+V_{\rm ee}\,,
\end{equation}
where
\begin{equation}
V_{\rm ext}(t)= -\frac{2}{\sqrt{1+x_1^2}} - \frac{2}{\sqrt{1+x_2^2}} + \mathcal{E}(t)(x_1+x_2)
\end{equation}
is the external potential with softened Coulomb interaction 
and the dipolar coupling to the external time-dependent field $\mathcal{E}(t)$
The electron-electron interaction is also described by a
soft-Coulomb function $V_{\rm ee} = \frac{1}{\sqrt{1+(x_1-x_2)^2}}$.
For the optimization we solve the equations discretized on a regular grid
of size $L=100$ a.u.\ and spacing $\Delta
x=0.2$~a.u.\ with 20~a.u.\ absorbing boundaries at the borders of the simulation box.
Results obtained with the optimized pulse were further converged
in a box of size $L=200$~a.u.\ with 70~a.u.\ absorbing boundaries. 
Time was discretized with a time step of $\Delta t = 0.025$~a.u.\ 
for a maximum propagation time of 1250~a.u.\ during optimization and 2250~a.u.\ for convergence.
The duration of the pump pulse was chosen to be $T_{\mathcal{P}}~=~800$~a.u.\ 
and the delay between pump and probe was set to $\tau=50$~a.u. for all the calculations. 
Finally the target region for optimization was chosen between 0.06~a.u.\ and 0.23~a.u.\ ($\approx$~200~nm and 800~nm).
We carried out optimizations at two different theory levels: exact (TDSE) and TDDFT 
with the adiabatic EXX functional (TDEXX) \cite{kummel-2003}.

\begin{figure}[htb!]
    \begin{center}
        \includegraphics[width=\columnwidth]{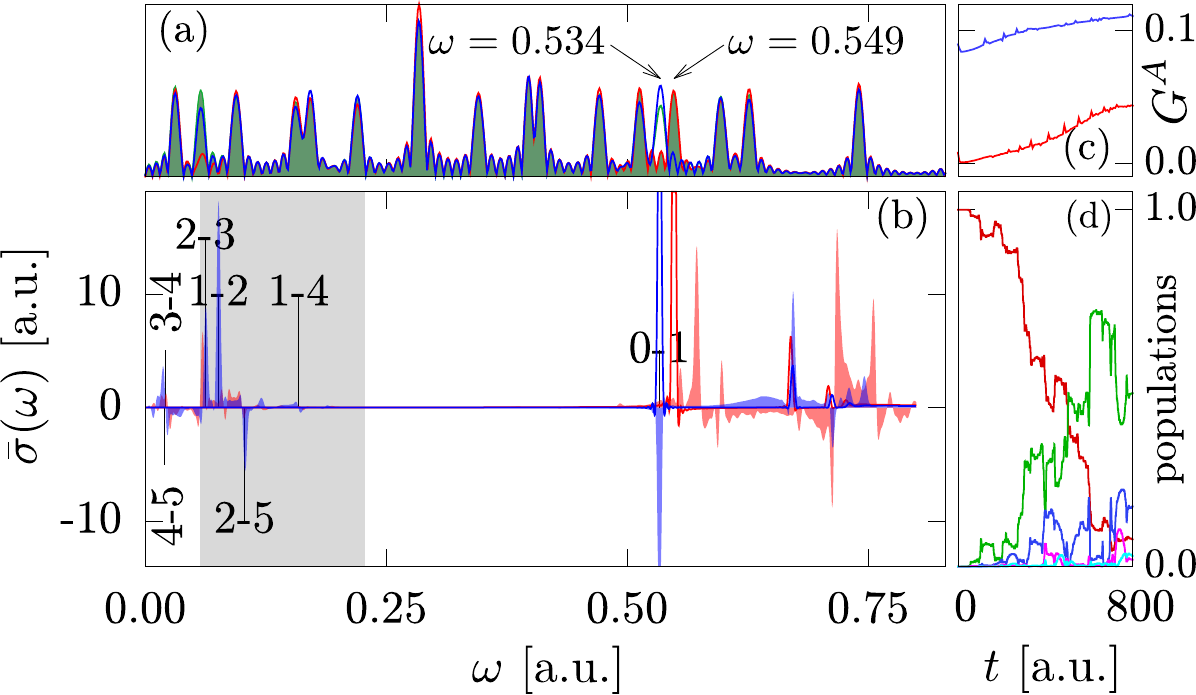}
    \end{center}
    \caption{Optimization of the absorption of one-dimensional Helium: TDSE vs.\ TDEXX. 
    (a) Power spectrum of the initial and optimized laser pulses: 
    The green line shows the initial laser used for the TDSE; the green shaded
    area shows the initial laser used for the TDEXX optimization (these two
    only differ by the indicated 
    peaks at $\omega=0.534$~a.u.\ and $\omega=0.549$~a.u.); the blue and red lines are
    the optimal pulses obtained when using TDSE and TDEXX, respectively.
    (b) Ground state (dashed line) and excited state (shaded) spectra of
    optimized one-dimensional Helium, in blue and red for the TDSE and TDEXX
    cases, respectively.
    The excited state transitions of the exact calculations are indicated.
    (c) The control function $G^A$ as a function of the number of iterations,
    also in blue and red for the TDSE and TDEXX, respectively.
    (d) The populations $|\langle \Psi(t)|\Psi_i\rangle|^2$ of the exact time propagation under the influence of the
    optimized pump pulse for the (red) ground state, (green) first excited state, (blue) second excited state, (pink)
    third excited state and (turquoise) fourth excited state.
    }
    \label{fig:he}
\end{figure}

Let us first focus on the optimization obtained by solving the exact TDSE
as illustrated in Figure~\ref{fig:he}\textcolor{blue}{(b)} where the ground state spectra 
are compared to the spectra of the systems excited by the optimized pump-pulses $\mathscr{E}_{100}(t)$ 
obtained after 100 iterations.
In the exact case, the search space was constructed from two
wave lengths $\lambda = 800$~nm and $\lambda = 1450$~nm and their first nine odd harmonics as shown in 
Figure~\ref{fig:he}\textcolor{blue}{(a)}. 
Optimization is achieved transferring population 
from the ground state (at $\epsilon_0=-2.238$~a.u.) into the first excited state (at $\epsilon_1=-1.705$) with the
help of the 9th harmonic of $\lambda=1450$~nm at
$\omega_{\mathcal{P}13}=0.534$~a.u.. Due to this population transfer, 
the peak at $\omega_{0\rightarrow 1}=0.533$~a.u.\ turns from positive to negative and peaks
coming from the first excited state ($\omega_{1\rightarrow 2}=0.076$~a.u.\ and $\omega_{1\rightarrow
4}=0.159$~a.u.) arise in the excited-state spectrum, where the peak at $\omega_{1\rightarrow 2}$ is located in the
visible part of the energy range. 
At the same time, population is transfered into the second excited
state ($\epsilon_2=-1.629$ a.u.), leading to e.g.\ the peaks at 
$\omega_{2\rightarrow 3}=0.062$~a.u.\ and $\omega_{2\rightarrow
5}=0.103$~a.u..
This interpretation is confirmed by the population analysis in Figure~\ref{fig:he}\textcolor{blue}{(d)}
where $|\langle\Psi(t)|\Phi_i\rangle|^2$ is plotted over time.
In particular it is apparent that at the end of the pump only $\approx$ 8\% of the electrons 
remain in the ground state, whereas the rest has been transfered into higher lying states 
thus explaining the appearance of the new peaks in the spectrum.  
To complete the picture in Figure~\ref{fig:he}\textcolor{blue}{(c)} we show the
evolution of the control function $G^A$ with the number of iterations. 
As can be seen, $G^A$ shows a steady increase during the optimization.

For the TDEXX case we adapted the search space by replacing the laser component
at the carrier frequency $\omega=0.534$~a.u. (in resonance with the first excitation in the TDSE case)
by a laser component with $\omega=0.549$~a.u., which is its TDEXX equivalent.
In our experience failing to meet this requirement resulted in poor optimizations.
The resulting optimization is shown in
Figure~\ref{fig:he}\textcolor{blue}{(b)}.
The results follow a trend similar to the exact case. However the TDEXX optimization is
smaller and the $1 \rightarrow 2$ peak present in TDSE seems to be missing. 
The difference between TDEXX and TDSE can be tracked down 
to a known problem of the adiabatic approximation in TDDFT. 
In particular, the lack of memory in the adiabatic 
approximation, causing a spurious time dependence of the exchange potential,
is responsible for the poor population transfer and the excess of asymmetric peaks in 
the spectrum~\cite{degiovannini-2013,Fuks:2015jp}.
This problem is further amplified by the ionization of the system, which 
results in an unphysical shift of the peaks to higher energies 
(compare the ground state and the excited state spectrum in Figure~\ref{fig:he}).
These effects, however, strongly depend on the fraction of the total density that gets driven
out of equilibrium and therefore become more dominant with decreasing size of the 
system -- with Helium being the worst case. 
In large molecules with many electrons we expect the error to be greatly reduced
(as has been empirically shown in studies of light induced charge transfer in organic photovoltaic
blends \cite{rozzi-2013, falke-2014}).

\begin{figure}%
    \begin{center}
      \includegraphics[width=\columnwidth]{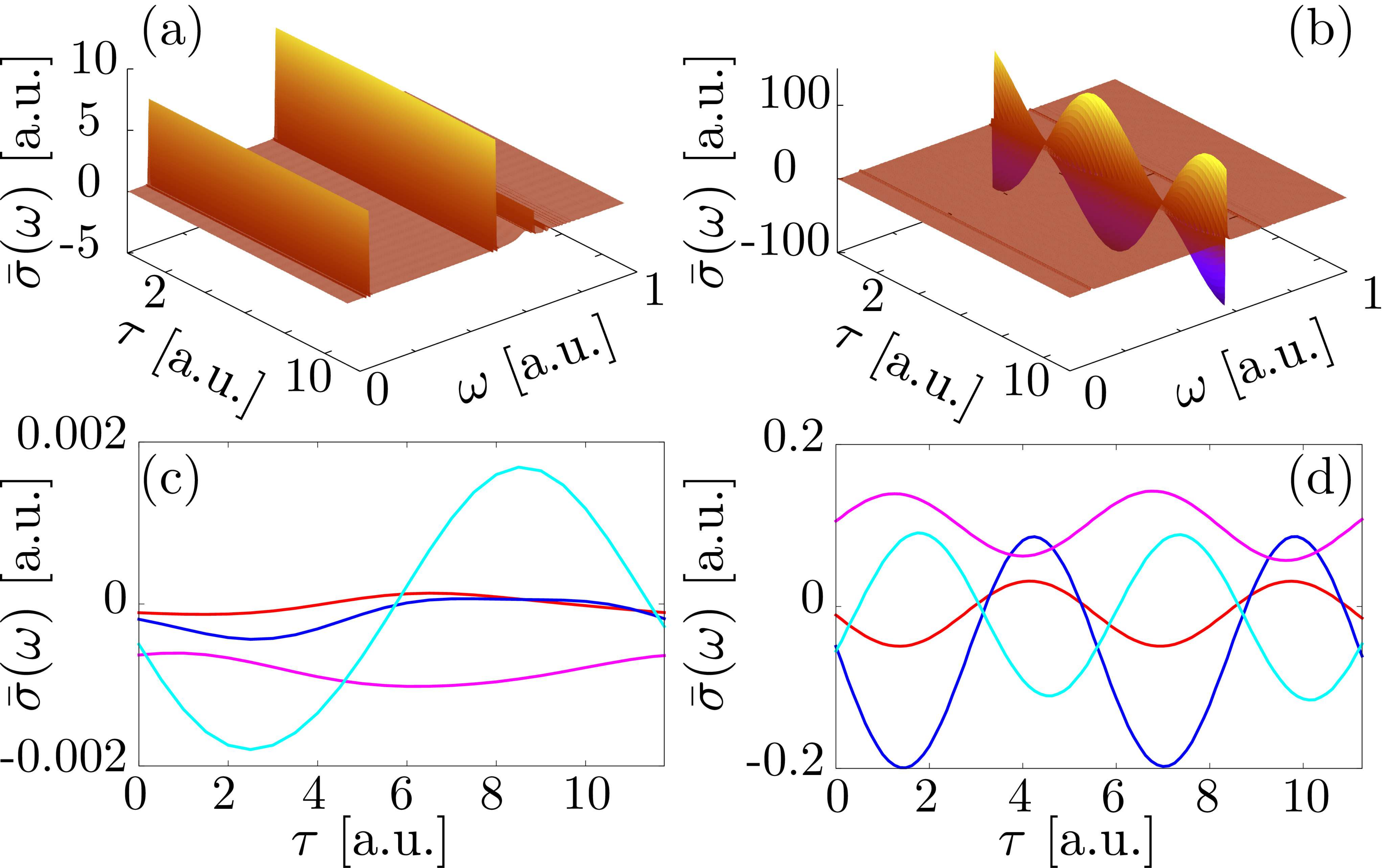}
    \end{center}
    \caption{(Top) Transient Absorption Spectrum of Helium after the excitation with a 45 cycle $\sin^2$ laser pulse of
    intensity $I=5.26 \cdot 10^{11} Wcm^{-2}$ with a carrier frequency resonant to the excitation energy from the ground
    to the first excited state for exact (a) and adiabatic EXX (b) with
    $\omega^{\rm exact}=0.534$~a.u. and $\omega^{\rm EXX}=0.549$~a.u.. Time-evolution of
    the absorption cross-section at selected energies $\omega_n = 0.2$ (red), 0.4 (blue), 0.6 (purple) 0.8 (turquoise) a.u.\ for
    exact (c) and adiabatic EXX (d). In the exact case the curve at 0.6~a.u.\ is offset by -0.1 for clarity.
    In all cases, the time interval $T=2\pi/(\epsilon_1-\epsilon_0)$ is shown.
    }
    \label{fig:he_tddft_time_resolved}
\end{figure}

A different perspective on the same problem can be obtained by comparing the time
evolution of an excited state spectrum in TDSE and TDEXX as shown in Figure~\ref{fig:he_tddft_time_resolved}.
The systems were excited by a 45 cycle $\sin^2$ laser pulse resonant with the excitation energy 
from the ground state into the first excited state. 
After the pulse the systems are in a superposition of these two states and
the spectra should contain time-dependent interference terms, which oscillate with the period time
$T=\frac{2\pi}{\epsilon_1-\epsilon_0}$, which is $T=11.76$~a.u.\ for TDSE and $T=11.26$~a.u.\ for TDEXX.
However, on the scale of Figure~\ref{fig:he_tddft_time_resolved}\textcolor{blue}{(a)} the TDSE spectrum hardly 
presents any oscillation. Therefore, in Figure~\ref{fig:he_tddft_time_resolved}\textcolor{blue}{(c)} we report 
cuts at $\omega_n$~=~0.2, 0.4, 0.6 and 0.8~a.u. From the figure it is apparent that, albeit with different phases, 
each cut presents oscillations with the expected period of $T=11.76$~a.u..
The TDEXX calculations, in Figure~\ref{fig:he_tddft_time_resolved}\textcolor{blue}{(b)} and \textcolor{blue}{(d)},
present a different picture. 
First of all, the amplitude of the oscillations is much larger than in the exact case and, second, the 
oscillations are two times faster than expected. 
We conclude, that the TDEXX description
seems to have a similar structure to the exact case, in the sense, that the energy difference of the involved states is
reflected in the periodicity of the oscillations of the spectrum. 
Nonetheless, there are major differences in the
behaviour, which is reflected in the factor of two in the periodicity. 

\subsubsection{Methane Dication}

Finally, we apply our scheme to a poly-atomic molecule: doubly-ionized Methane, $\mathrm{CH}_4^{+2}$.
The goal here is to design a laser capable to turn this molecule, transparent 
in nature, visible. 
To this end we used the same strategy as we did before for Helium, namely 
we optimize the laser on a small simulation box and then converge the results with the 
optimized laser on a larger box. 
During the optimization routine the simulation box has a radius of $R = 15$~a.u., including 5~a.u.\
absorbing boundaries while the results are converged in a box of $R = 30$~a.u.\ with 15~a.u.\ absorbing boundaries.
We discretize the TDDFT equations on a three-dimensional grid with a spacing of $\Delta x =0.3$~a.u..
The reason for this box choice is the fact that the computational 
costs of three-dimensional calculations scale with the third power of 
the simulation box radius.
The maximum propagation time is 850~a.u.\ during the optimization and 1600~a.u.\ for convergence. In all cases, the 
pump duration was 600~a.u., the time step was $\Delta t = 0.04$~a.u.\ and the delay was $\tau=0$~a.u..

The optimization region was chosen as the interval between 
0.057~a.u.\ and 0.139~a.u.~(328 and 750~nm) and 
in order to discourage the algorithm from exciting too many electrons into
the continuum, we used the target functional $G^B_{\tau}$ \eqref{eq:epsilon_b}, 
which includes an exponential ``penalty'' for ionization.

To obtain a good description of states close to the ionization threshold, 
we employed the average density self-interaction corrected (ADSIC) LDA
functional \cite{legrand-2002}, which is asymptotically correct. 

\begin{figure}[htb!]
    \begin{center}
        \includegraphics[width=\columnwidth]{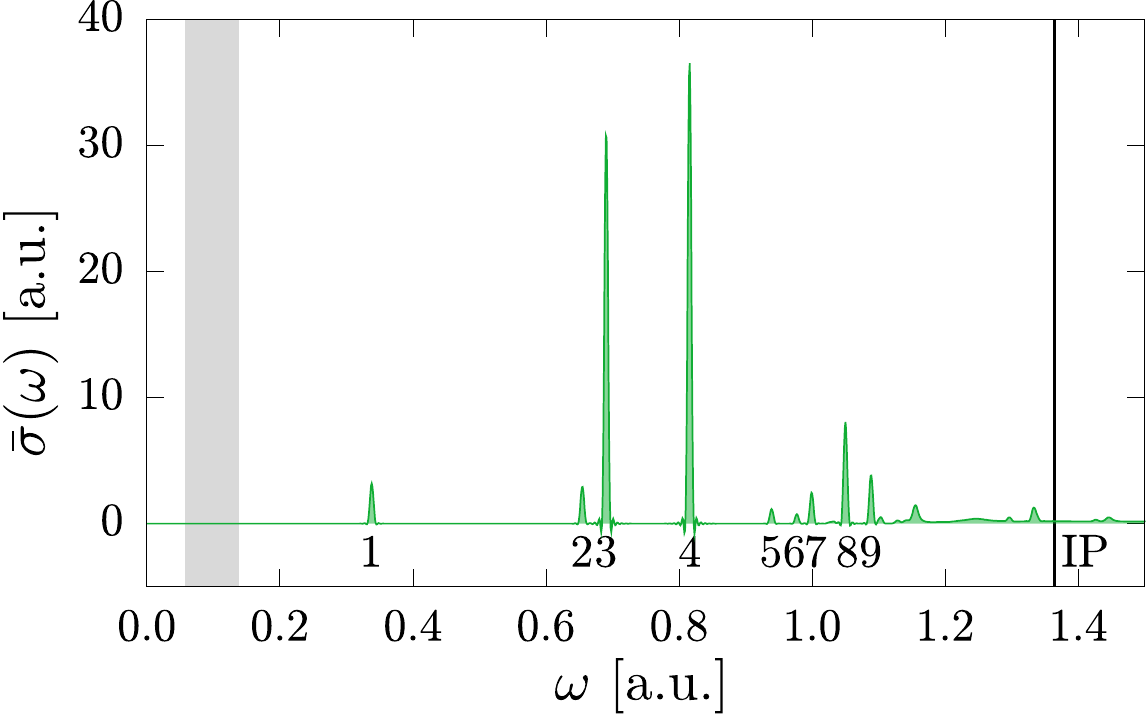}
    \end{center}
    \caption{Ground state spectrum of doubly-ionized Methane CH$_4^{2+}$. Peaks are numbered for later reference (as
    discussed in the text). The
    shaded grey area marks the optimization range.}
    \label{fig:ch4_2plus_gs}
\end{figure}

The inclusion of resonant frequencies in the search space is a good practice 
that facilitates transitions between eigenstates and enables the optimization algorithm to
populate excited eigenstates.
These molecular excitation frequencies can easily be obtained from 
the ground-state spectrum reported in Figure~\ref{fig:ch4_2plus_gs}.
By populating the correct
eigenstates, the system might absorb in the visible region: consider
two eigenstates with energies $\epsilon_h$ and $\epsilon_T$, 
that differ by an energy in the visible: 0.057~a.u.~$\leq \epsilon_h-\epsilon_T\leq$~0.139~a.u.. 
By exciting the system into the lower ``target'' state $\epsilon_T$ 
one might obtain transition peaks in the visible, due to the transition to the
higher one. 
Note, however, that this 
fact is not guaranteed since the transition might be dipole forbidden. We
cannot rule out this possibility since our groundstate linear-response TDDFT
calculation does not provide this information. 
The ground state spectrum shows, that the first possible target
state is $\epsilon_3$. The energy difference between
$\epsilon_3=0.690$~a.u.\ and $\epsilon_4=0.816$~a.u.\ is $\omega_{3\rightarrow 4}=0.126$~a.u.\ 
and lies -- with 362 nm -- at the red end of the visible spectrum. 
Also $\epsilon_4$ provides a transition in the visible range -- into $\epsilon_5=0.938$~a.u.\ 
with $\omega_{4\rightarrow 5}=0.122$~a.u.\ = 373 nm. 
Starting from $\epsilon_5$, the  states have even more than one transition in the
visible. We must therefore choose a frequency search space, 
that allows the construction of a pump pulse, that excites electrons
from the ground state into $\epsilon_3$ and higher lying states either directly or by successive excitations.

Here we present results for two possible search spaces.
The first search space includes frequencies that are either resonant 
to the ground state excitation energies $\epsilon_n$, or to excited state excitations $\epsilon_m-\epsilon_n$.
To avoid ionization, all carrier frequencies are smaller than $\epsilon_7=1.0$ a.u..
The second frequency search space was designed using the ionization potential $I_P$ 
of the system (which is equal to minus the 
energy $\epsilon_{H}$ of the highest occupied KS orbital $\varphi_H$) and the energy differences $I_P-\epsilon_n$. 
One frequency is $I_P$ itself, while all others are resonant with 
relaxations bringing down states at that ionization threshold
to bound excited states ($\epsilon_3$ to $\epsilon_8$).
The idea is that the system could be excited into the ionization threshold,
and then relax into one of the target states. 
The laser frequencies, that were included in the search spaces and the corresponding resonances are summarized
in Appendix~\ref{sec:appendix_laserfreqs}.
\begin{figure}[htb!]
  \begin{center}
      \includegraphics[width=\columnwidth]{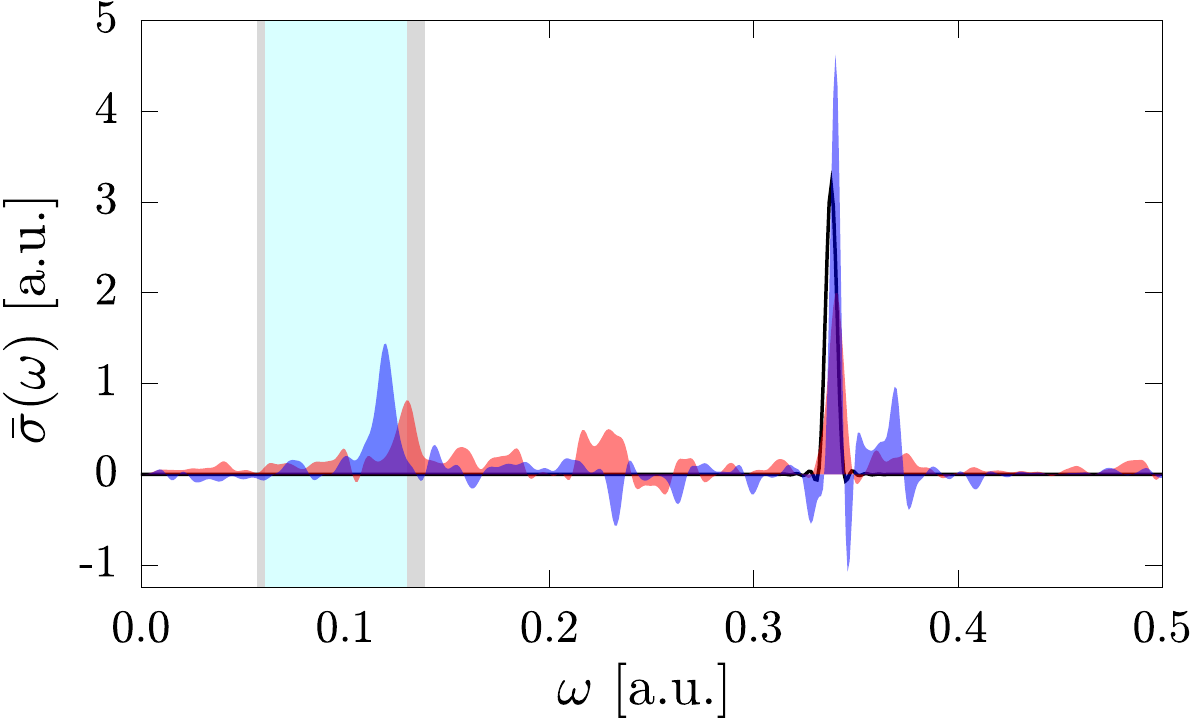}
  \end{center}
    \caption{
    Ground state (black line) and excited state (shaded) spectra of 
    doubly-ionized Methane (CH$_4^{2+}$) 
    for two different pump pulses $\mathscr{E}^I$ (red) and $\mathscr{E}^{II}$ (blue). 
    The grey shaded box marks the optimization area, the blue shaded area the visible region of the
    spectrum.
    }
  \label{fig:ch4_2plus}
\end{figure}

The optimized spectra are shown in Figure~\ref{fig:ch4_2plus}.
It can be seen that both search spaces include optimal lasers that cause the molecule to loose its transparency and 
absorb in the visible. The achieved opacity can be quantified in terms of the control function $G^A$ \eqref{eq:epsilon_a} 
being the integral over the absorption spectrum in the visible range of the spectrum. Comparing the opacity achieved 
in search space I ($G^A_I=0.017$) with the one achieved in search space II ($G^A_{II}=0.020$), we conclude that 
search space II is better suited for the pursued optimization. Thus, including energy levels at the ionization
threshold in the search space might be a useful strategy in further optimizations. 

\section{Conclusions} %
\label{sec:conclusions}

In this work, we assessed the possibility of using tailored pumps in order to
enhance some given features of the probe absorption -- for example, the
absorption in the visible range of otherwise transparent samples. 
We first detailed a theoretical analysis of the
non-equilibrium response function in this context, aided by one simple
numerical model of the Hydrogen atom. Then, we investigated the
feasibility of using TDDFT theory as a means to
implement, theoretically, this absorption-optimization idea, for more complex
atoms or molecules.

The theoretical analysis of the response function can be done by writing it in
a generalized form of the Lehmann representation, valid for systems that have
been pumped out of equilibrium by a first pulse, and whose response to a probe
pulse (in our case, assumed non-overlapping with the first one) needs to be
studied and manipulated. The peaks of this response functions are always fixed
to the differences in the system energies, but their strength and shape varies
depending on the pump shape, and on the pump-probe delay. Furthermore, the
response function is a sum of a stationary part (the only one present if the
pumped state is itself a stationary state), and a time-dependent,
oscillatory term, caused by interferences between the populated eigenstates.

We then used this dependence of the non-equilibrium response with respect to the pump pulse
shape to manipulate it by means of QOCT. We demonstrated the idea first with a
small model, that could be treated analytically. This could be a viable
alternative for larger systems, if they can be reduced to few-level
models. However, for full generality we also showed how QOCT can be combined
with TDDFT. We showed how this avenue is tractable, but we also highlighted
the key numerical difficulties and theoretical challenges. For this purpose,
we performed first calculations on a model for the Helium atom that could be
solved both exactly with the TDSE equation, and with TDDFT within the
adiabatic EXX approximation. Then we concluded with simulations of the
methane dication.

From our results we conclude that the proposed idea could be brought to the
laboratory: tailored pump pulses can excite systems into light-absorbing states.
Theoretically, the scalability of TDDFT could in principle permit studying
these processes for larger systems. However, our results have also highlighted
the severe numerical and theoretical difficulties posed by the problem: large-scale
non-equilibrium quantum dynamics are cumbersome, even with TDDFT, and moreover
the shortcomings of state-of-the-art TDDFT functionals may still be
serious for these out-of-equilibrium situations. 
Our findings confirm recent investigations about the consequences of these shortcomings 
for the use of coherent control schemes~\cite{raghunathan-2011, raghunathan-2012}.

\section*{Acknowledgements} %

We acknowledge financial support from the European Research Council (ERC-2010-AdG-267374), Spanish grant
(FIS2013-46159-C3-1-P), Grupos Consolidados (IT578-13), and AFOSR Grant No. FA2386-15-1-0006 AOARD 144088,
H2020-NMP-2014 project MOSTOPHOS, GA no. SEP-210187476 and COST Action MP1306 (EUSpec).
Computational time was granted by BSC Red Espanola de Supercomputacion.

\appendix

\section{Quantum Optimal Control Equations}\label{sec:qoct_gradientbased}

For completeness, we derive here the equations for
the computation of the gradient of a target functional designed to optimize
the response of a system.
In general, the equation for the gradient provided by QOCT is given by:
\begin{subequations}
\label{eq:equations_of_control}
\begin{align} 
\nonumber
\nabla_\mathbf{c}& G[\mathbf{c}] = \left.\nabla_\mathbf{c}
  F[\Psi,\mathscr{E}_\mathbf{c}]\right|_{\Psi = \Psi[\mathscr{E}_\mathbf{c}]}  +
\\ 
&  2 \Im \int_0^T\!\!\!{\rm d}t\; \langle \chi[\mathscr{E}_\mathbf{c}](t)
 \vert \nabla_{\mathbf{c}}
\hat{H}[\mathscr{E}_\mathbf{c},t] \vert \Psi[\mathscr{E}_\mathbf{c}](t)\rangle\,
\label{eq:qoct-gradient}
\end{align}

Note that a new ``wave function'', $\chi[\mathscr{E}]$, has been introduced;
it is given by the solution of:
\begin{eqnarray}
\label{eq:lambda-1}
i\frac{\partial \chi[\mathscr{E}]}{\partial t} (x,t)  & = & \hat{H}^{\dagger}[\mathscr{E},t]
\chi[\mathscr{E}](x,t) 
\\
\label{eq:lambda-2}
\chi[\mathscr{E}](x,T) & = & \frac{\delta J_1^T}{\delta \Psi^*[\mathscr{E}](x,T)} \,.
\end{eqnarray}
\end{subequations}
This is similar to the original Schr{\"{o}}dinger equation~\eqref{eq:schroedinger_control}, although
the initial condition is
given at the final time $t=T$, which implies it must be propagated
\emph{backwards}.
For a detailed derivation of
Eqs.~\eqref{eq:equations_of_control} we
refer the reader to Refs.~\cite{peirce-1988, kosloff-1989, ohtsuki-2004, serban-2005}. 

The computation of the gradient or functional derivative of $G$,
therefore, requires $\Psi[\mathscr{E}]$ and $\chi[\mathscr{E}]$, which are obtained by
first propagating Eq.~\eqref{eq:schroedinger-1} forwards, and then
Eq.~\eqref{eq:lambda-1} backwards. The maxima of $G$ are found at the
critical points $\nabla_\mathbf{c} G[\mathscr{E}_\mathbf{c}] = 0$.

We may now apply these general equations for
a target functional designed to optimize
the response of a system after the excitation by a pump pulse. 
This setup is consistent with the non-overlapping regime described
in Section~\ref{sec:spec_ex}, where the Hamiltonian that governs the system, once that the pump has passed
($t \ge T$) has the form \eqref{eq:hamiltonian_nonoverlapping}.
If, at time $t=T$, the system has been driven to the state
$\vert\Psi(T)\rangle$, the response function for the perturbation at
later times is given by \eqref{eq:retarded_response_function_recast}
and the first-order response of the system is given by \eqref{eq:dipole_response}.

The key point is the definition of a target: for example, let us assume, that we
wish to enhance the
reaction of the system at a given frequency to a sudden perturbation at 
the end of the pump $F(t') = \delta(t-T)$. 
As seen in \eqref{eq:first_order_response_deltafunction}, the time-dependent dipole-dipole 
response is then directly given by the response function
$\chi_{\hat{D},\hat{D}}\left[\mathscr{E}\right](t,T) = D^{(1)}\left[\mathscr{E}, \delta_{T}\right](t) $
and its Fourier transform by
\begin{eqnarray}\nonumber
\chi_{\hat{D},\hat{D}}\left[\mathscr{E}\right](\omega,T) 
    &=& D^{(1)}\left[\mathscr{E}, \delta_{T}\right](\omega) \\
    &=& \int_{T}^{\infty}\!\!{\rm d t'}\;e^{-i\omega t'}D^{(1)}\left[\mathscr{E}, \delta_{T}\right](t') \, . 
    \label{eq:response_ft}
\end{eqnarray}
It can be easily seen that the problem fits into
the framework discussed above, i.e.\ the target functional is
given by the expectation value of some operator:
\begin{align}
\nonumber
&J_1^T[\Psi(T)] = -i
\\
&\langle \Psi(T)\vert
\int_{T}^{\infty}\!\!{\rm d t'}\;e^{-i\omega
  t' }\left[e^{i( t'-T)\mathscr{H}}\hat{D}
e^{-i(t'-T)\mathscr{H}},\hat{D}\right]
\vert\Psi(T)\rangle\, .
\end{align}
The equation for the gradient is therefore Eq.~\eqref{eq:qoct-gradient}; 
which must be completed with the equation of motion for the
co-state, Eq.~\eqref{eq:lambda-1}, and, in particular, with its boundary
condition~\eqref{eq:lambda-2} at time $t=T$: this is the only one that in fact depends on the
definition of the target operator:
\begin{align}
\nonumber
&\vert\chi(T)\rangle = -i
\\
&\int_{T}^{\infty}\!\!{\rm dt'}\;e^{-i\omega
  t'}\left[e^{i(t'-T)\mathscr{H}}\hat{D} e^{-i(t'-T)\mathscr{H}},\hat{D}\right]
\vert\Psi(T)\rangle\,.
\end{align}
Similar formulas can be obtained for more general definitions of the target
functional in terms of the response $D^{(1)}(\omega)$, and for more general probe
functions. In all cases the computational difficulties associated to the
computation of this boundary condition are similar, and are
considerable. By inspecting the previous formula, it can be learnt that
various time-propagations of the wave functions, forwards and backwards, are
required.  These difficulties are even larger if the scheme is formulated
within TDDFT -- in the previous derivation we have used the exact
many-electron wave functions. In consequence, we decided to employ, for this
type of optimizations, gradient-free algorithms, such as the Simplex-Downhill algorithm 
that we describe in next section.

\section{Derivation of the Control Equations for Three Level
  Systems}\label{sec:analyticalcontrol}

For the three-levels model described at the end of Section~\ref{sec:qoct},
we will start by considering a simpler situation in which the field envelopes
are constant, i.e.:
\begin{equation}
    \mathscr{E}(t)=\varepsilon_1\cos(\omega_1 t+\varphi_1) + \varepsilon_2\cos(\omega_2 t + \varphi_2) \, .
\end{equation}
If the two carrier frequencies are sufficiently close to the transition
frequencies $\omega_{ab}$ and $\omega_{bc}$, one can apply the rotating wave
approximation (RWA), as it is done in the theory of 
Rabi oscillations. In fact, we choose the carrier frequencies to be equal to the transition energies. 
In addition, we also assume, that the laser frequencies are sufficiently well separated in energy to apply the RWA a second time:
\begin{subequations}
\begin{eqnarray}
    |\omega_1+\omega_2| &\gg& 0 \, ,  \\
    |\omega_1-\omega_2| &\gg& 0 \, .
\end{eqnarray}
\end{subequations}
Assuming the validity of the RWA mentioned in Section~\ref{sec:qoct}, the 
solution of the resulting differential equations with the initial conditions $a(0)=1$, $b(0)=c(0)=0$, 
leads to the following time-evolution of the coefficients:
\begin{subequations}
    \label{eq:rabi_solutions}
\begin{eqnarray}
   a(t) &=& \cos(\bar\Omega/2 t) \, , \\
   b(t) &=& \frac{d_{ab}\varepsilon_1e^{-i(\varphi_1-\pi)}}{\sqrt{
   (d_{ab}\varepsilon_1)^2+(d_{ac}\varepsilon_2)^2}}\sin\left(\bar\Omega/2 t\right) \, , \\
   c(t) &=& \frac{d_{ac}\varepsilon_2e^{-i(\varphi_2-\pi)}}{\sqrt{
   (d_{ab}\varepsilon_1)^2+(d_{ac}\varepsilon_2)^2}}\sin\left(\bar\Omega/2 t\right) \, ,
\end{eqnarray}
\end{subequations}
with the Rabi-frequency
\begin{eqnarray}
    \bar\Omega = \sqrt{(d_{ab}\varepsilon_1)^2+(d_{ac}\varepsilon_2)^2} \, .
    \label{eq:rabi_frequency_3level}
\end{eqnarray}
Note that:
\begin{enumerate}
    \item The Rabi-frequency is the Pythagorean mean 
            of the Rabi-frequencies of the single transitions:
            $\bar \Omega = \sqrt{\bar\Omega_{ab}^2+\bar\Omega_{ac}^2}$. Consequently, it is larger 
            than each of those single frequencies.
    \item The maximum populations of the excited states depend only on the ratio of the Rabi-frequencies belonging to
        the respective transitions $\frac{|b(t)|^2}{|c(t)|^2}=\frac{\bar\Omega_{ab}^2}{\bar\Omega_{ac}^2}$.
    \item The relative phases of the expansion coefficients depend on the phases of the applied lasers.
\end{enumerate}

The target state is defined in Eq.~\eqref{eq:hcontrol_target}: the goal is to
find a laser pulse that drives the system from the state $|\Psi(t=0)\rangle=|\Phi_a\rangle$ into this target state
within the time $T$. The evolution of the time-dependent wave function is
given by:
\begin{align}\nonumber
   |\Psi(t)\rangle &= \cos(\bar\Omega/2 t) |\Phi_a\rangle \\\nonumber
   & + \frac{d_{ab}\varepsilon_1\sin\left(\bar\Omega/2 t\right)}{\sqrt{ (d_{ab}\varepsilon_1)^2+(d_{ac}\varepsilon_2)^2}}
        e^{-i(\varphi_1-\pi+\omega_{ba}t)}|\Phi_b\rangle \\
   & + \frac{d_{ac}\varepsilon_2\sin\left(\bar\Omega/2 t\right)}{\sqrt{ (d_{ab}\varepsilon_1)^2+(d_{ac}\varepsilon_2)^2}}
        e^{-i(\varphi_2-\pi+\omega_{ca}t)}|\Phi_c\rangle \, .
\end{align}
The condition $\vert\langle\Phi_a\vert\bar{\Psi}\rangle\vert^2 = 1$
leads to two sets of equations: one connecting the laser amplitudes $\varepsilon_1$ and
$\varepsilon_2$ to the populations $|\alpha|^2$, $|\beta|^2$ and $|\gamma|^2$
\begin{subequations}
    \label{eq:hcontrol_populations}
\begin{eqnarray}
   |\alpha| &=& \cos(\bar\Omega/2 T) \, , \\
   |\beta|  &=& \frac{d_{ab}\varepsilon_1}{\sqrt{ (d_{ab}\varepsilon_1)^2+(d_{ac}\varepsilon_2)^2}}\sin\left(\bar\Omega/2 T\right) \, , \\
   |\gamma| &=& \frac{d_{ac}\varepsilon_2}{\sqrt{ (d_{ab}\varepsilon_1)^2+(d_{ac}\varepsilon_2)^2}}\sin\left(\bar\Omega/2 T\right) \, , 
\end{eqnarray}
\end{subequations}
and the other one connecting the laser phases to the relative phases $\varphi_{\beta}$ and
$\varphi_{\gamma}$ of the wave function
\begin{subequations}
\begin{eqnarray}
    \varphi_{\beta} &=&  \varphi_1-\pi+\omega_{ba}T \, , \\
    \varphi_{\gamma} &=& \varphi_2-\pi+\omega_{ca}T \, .
\end{eqnarray}
\end{subequations}
Solving these sets of equations, we find the following amplitudes as one example of a control laser
\begin{eqnarray}
\varepsilon_1 & = &
\frac{2}{T}\frac{\arccos(|\alpha|)}{\sin(\arccos(|\alpha|)} 
\frac{|\beta|}{d_{ab}}\, ,
\\
\varepsilon_2 & = &
\frac{2}{T}\frac{\arccos(|\alpha|)}{\sin(\arccos(|\alpha|)} 
\frac{|\gamma|}{d_{ac}}\,.
\end{eqnarray}
Note, however, that the solutions are not unique: other sets
$\{(2n+1)\varepsilon_1, (2n+1)\varepsilon_2\}$ fulfill the equations above. 
These solutions represent lasers that lead to an
evolution of the coefficients that covers $(n+1)$ complete Rabi cycles within
the time $T$.

In practice one is often interested in pulses with time-dependent envelope
functions, such as the ones discussed in Section~\ref{sec:qoct}, in which
the pulses have a $\sin^2$ envelope with a period of $2T$. The problem can be
solved in an analogue manner; the solutions for
the amplitudes were already given in Section~\ref{sec:qoct}. In this case, the
evolution of the coefficients is given by:
 \begin{subequations}
     \label{eq:hcontrol_populations_td}
\begin{eqnarray}
    a(t) &=& \cos\left(\frac{\int_0^t\tilde\Omega(t')dt'}{2}\right)\, , \\
    b(t) &=& \frac{d_{ab}\varepsilon_1e^{-i(\varphi_1-\pi)}}{\sqrt{
    (d_{ab}\varepsilon_1)^2+(d_{ac}\varepsilon_2)^2}}\sin\int_0^t\frac{\tilde\Omega(t')dt'}{2}\, ,\\
    c(t) &=& \frac{d_{ac}\varepsilon_2e^{-i(\varphi_2-\pi)}}{\sqrt{
    (d_{ab}\varepsilon_1)^2+(d_{ac}\varepsilon_2)^2}}\sin\int_0^t\frac{\tilde\Omega(t')dt'}{2}\,,
\end{eqnarray}
\end{subequations}
where the ``time-dependent Rabi-frequency'' $\tilde\Omega(t)$ is given by:
 \begin{equation}
     \tilde\Omega(t)=\sqrt{(d_{ab}\tilde\varepsilon_1(t))^2+(d_{ac}\tilde\varepsilon_2(t))^2} =
     2\bar\Omega\sin^2\left(\pi\frac{t}{T}\right) \, .
 \end{equation}
and integrates as:
\begin{equation}
    \int_0^t \frac{\tilde\Omega(t')dt'}{2} = \left(\frac{1}{2}t - \frac{T}{4\pi}\sin\left(2\pi\frac{t}{T}\right)\right)\bar\Omega\, .
\end{equation}

\section{Laser Frequencies Used in the Optimization of
  Methane}\label{sec:appendix_laserfreqs}

The laser frequencies (in a.u.) and the corresponding resonances 
of the search spaces of the optimization of CH$_4^{2+}$. The nomenclature
follows the one in Fig.~\ref{fig:ch4_2plus_gs}: $\epsilon_H$ is minus the energy of the highest occupied KS state.
$\omega_3^I$ is the average of $\epsilon_1=0.337$~a.u.\ and $\epsilon_3-\epsilon_1=0.353$~a.u.. 
Since the frequencies are broadened by the finite pulse duration, 
$\omega_3^I$ covers both resonances.

    \centering
    \begin{tabular}{c|c|c|c|c} 
        \multicolumn{5}{c}{Search Space I}\\\hline\hline
        $\omega_1^I$ & $\omega_2^I$ &$\omega_3^I$ &$\omega_4^I$ &$\omega_5^I$  \\
        0.122       & 0.248     & 0.345             & 0.479 & 0.601 \\
        $\epsilon_4-\epsilon_5$     & $\epsilon_3-\epsilon_5$   & $\epsilon_1$
        / $\epsilon_1-\epsilon_3$      & $\epsilon_1-\epsilon_4$   &
        $\epsilon_1-\epsilon_5$ \\
        \hline
        $\omega_6^I$ &$\omega_7^I$ & $\omega_8^I$ &$\omega_9^I$ &$\omega_{10}^I$ \\
        0.654 & 0.690 & 0.816 & 0.938 & 1.000\\
        $\epsilon_2$    & $\epsilon_3$ & $\epsilon_4$ & $\epsilon_5$ &
        $\epsilon_7$  \\\hline
    \end{tabular}
    \begin{tabular}{c|c|c|c|c|c|c}
        \multicolumn{7}{c}{}\\
        \multicolumn{7}{c}{Search Space II}\\\hline\hline
        $\omega^{II}_1$ & $\omega^{II}_2$ & $\omega^{II}_3$ & $\omega^{II}_4$ & $\omega^{II}_5$ & $\omega^{II}_6$ & $\omega^{II}_7$ \\
        0.311 & 0.364 & 0.386 & 0.426 & 0.548 & 0.674 & 1.364 \\
        $\epsilon_3 - I_P$ & $\epsilon_4-I_P$ & $\epsilon_5-I_P$ & $\epsilon_6-I_P$ & $\epsilon_7-I_P$ &
        $\epsilon_8- I_P$ & $I_P$ 
        \\\hline
    \end{tabular}

\clearpage
\bibliography{walkenhorst}

\end{document}